\documentclass[12pt]{article}
\usepackage{amsmath,amssymb,epsfig}

\newcommand{\beq}{\begin{equation}}
\newcommand{\eeq}{\end{equation}}
\newcommand{\ba}{\begin{array}}
\newcommand{\ea}{\end{array}}
\newcommand{\bea}{\begin{eqnarray}}
\newcommand{\eea}{\end{eqnarray}}
\newcommand{\bean}{\begin{eqnarray*}}
\newcommand{\eean}{\end{eqnarray*}}
\newcommand{\eref}[1]{(\ref{#1})}

\newcommand{\comment}[1]{}

\newcommand{\CN}{{\cal N}}

\newcommand{\IC}{\mathbb{C}}

\newcommand{\IZ}{\mathbb{Z}}
\newcommand{\f}{{\cal F}^{\flat}}
\newcommand{\firr}[1]{{}^{{\rm Irr}}\!{\cal F}^{\flat}_{#1}}

\newcommand{\setall}{\setcounter{equation}{0}
        \setcounter{theorem}{0}}

\newcommand{\tmat}[1]{{\tiny \left(\begin{matrix} #1 \end{matrix}\right)}}

\begin{document}
\setcounter{page}{0}
\begin{titlepage}
\titlepage
\begin{flushright}
Bicocca-FT-08-16\\
CERN-PH-TH/2008-200\\
Imperial/TP/08/AH/09 \\
\end{flushright}
\begin{center}
\LARGE{\Huge Master Space, Hilbert Series\\}
\LARGE{\Huge and Seiberg Duality }
\end{center}
\vskip 0.5cm \centerline{{\bf \large Davide Forcella$^{a,b}$ \footnote{\tt davide.forcella@cern.ch}
    Amihay Hanany$^{c}$ \footnote{\tt a.hanany@imperial.ac.uk},  
Alberto Zaffaroni$^{d}$ \footnote{\tt alberto.zaffaroni@mib.infn.it}}}
\medskip
\footnotesize{

\begin{center}
$^a$ International School for Advanced Studies (SISSA/ISAS) 
\& INFN-Sezione di Trieste, via Beirut 2, I-34014, Trieste, Italy\\
\medskip
$^b$ PH-TH Division, CERN 
CH-1211 Geneva 23, Switzerland\\
\medskip
$^c$ Theoretical Physics Group, Blackett Laboratory, 
Imperial College, London SW7 2AZ, U.K. \\
\medskip
$^d$ Universit\`{a} di Milano-Bicocca and INFN, sezione di Milano-Bicocca, Piazza della Scienza, 3; I-20126 Milano, Italy
\end{center}}

\bigskip

\begin{abstract}

We analyze the  action of Toric (Seiberg) duality on  the combined mesonic and baryonic moduli space of quiver gauge theories obtained from D3 branes at Calabi-Yau singularities. We analyze in particular the structure of the master space, the complete moduli space for one brane, for different toric phases of a given singularity. We show that the Hilbert Series for the largest component of the master space of different phases is the same, when refined with all the non anomalous charges. This reflects the fact that the quiver gauge theories associated with different phases are related by Seiberg duality when the number of branes is greater than one.

\end{abstract}

\vfill
\begin{flushleft}
{\today}\\
\end{flushleft}
\end{titlepage}

\newpage

\tableofcontents

\section{Introduction}\setall

The moduli space and the BPS operators are very important concepts for $\mathcal{N}=1$ supersymmetric field theories. These have been recently analyzed, in the context of D3 branes at singularities, using the concepts of Master space, Hilbert series and Plethystic exponential in \cite{Benvenuti:2006qr, Feng:2007ur, Butti:2006au,Hanany:2006uc, Forcella:2007wk,Forcella:2007ps,Forcella:2008bb,Forcella:2008eh,Butti:2007jv}. The moduli space and the chiral ring are sometimes modified by non perturbative dynamics. It is interesting to see if we can get some information about this dynamics using the Master space and the Hilbert series \footnote{One example in this direction is the relation discovered in \cite{Forcella:2008au} between stringy instantons and BPS operators.}. In this paper we focus on Seiberg Duality. This is a quantum field theory duality that it is realized as Toric Duality in the setup of D3 branes at singularities.
Toric Duality was discovered in \cite{Feng:2000mi}, analyzed in \cite{Feng:2001xr,Feng:2002zw, Feng:2002fv} and identified as Seiberg duality in \cite{Beasley:2001zp,Feng:2001bn}. It corresponds to a situation in which one singular CY manifold has more than one quiver gauge theory that has this manifold as its mesonic moduli space of vacua. Given a CY singularity, $X$ there are in fact an infinite number of gauge theories that have $X$ as their mesonic moduli space of vacua and there are several studies of this in the literature. See for example the use of Picard Lefshetz transformations introduced in this context in \cite{Hanany:2001py,Feng:2002kk} to generate the corresponding duality trees \cite{Franco:2002mu}. There is however a special subset of quiver gauge theories that are the focus of the present paper. If the singularity $X$ is toric, then this special subset is characterized by having the ranks of all the gauge groups equal, and each field in the quiver appears pr!
 ecisely twice in the superpotential. Such a quiver theory is called a toric phase and the singularity $X$ may have more than one toric phase. The number of toric phases is in fact a finite number and it is an interesting problem to count this number for a given singularity $X$. The analysis of toric phases for certain classes of quiver gauge theories called the $Y^{pq}$ theories was done in \cite{Benvenuti:2004wx} and was found to have an exponential growth in $p$ for small values of $q$.

From a field theory perspective, for theories which are not necessarily living on D3 branes at singularities, Seiberg dual theories have the same moduli space and the same spectrum of chiral operators. This nicely agrees with our understanding of the mesonic moduli space for the theories on D3 branes at singularities. It is $X$ in the abelian case, and ${\rm Sym}^N X$ in the non-abelian case, for all toric phases. This leaves a question on the baryonic moduli space or alternatively the combined mesonic and baryonic moduli space for this class of theories. This is the subject of investigation in the present paper.

In this paper we look at a collection of examples, $X$, which have more than one toric phase and we will study their full moduli space, including baryonic directions.  A systematic investigation of the full spectrum of chiral operators has been carried out in the series of papers \cite{ Butti:2006au,Forcella:2007wk,Forcella:2007ps,Forcella:2008bb,Forcella:2008eh,Butti:2007jv}. A crucial ingredient in the analysis is the concept of the {\bf master space}, which is the complete moduli space for one brane.  In the case of one brane all gauge groups, being abelian, decouple in the IR and we are left with a theory of chiral multiplets and no gauge interactions (these still play a crucial role and turn into global baryonic symmetries). The master space is generically reducible into different components that have a variety of structures and a variety of dimensions. There is one large component called the coherent component and there are a number of other lower dimensional (generically linear) components. Here linear means copies of some $\IC^{l}$ for some value of $l$. In this paper we study the action of toric duality on the master space and we compute the Hilbert series for each toric phase.  It turns out that the coherent component of the master spaces of different phases are in general not isomorphic and the number of lower dimensional components differ between the phases. Furthermore, the fully refined Hilbert series for the coherent component, written in terms of fugacities for all the classical global symmetries of the theory, are not equal for different phases. This is a somewhat disappointing result but there is however a point of light. Some of the global symmetries are anomalous and some are anomaly free. We discover that the Hilbert series of the coherent component restricted to the set of non-anomalous charges is the same in all phases. In particular, for non-chiral theories, where all the abelian symmetries are non anomalous, we conclude that the coherent components of different phases are isomorphic. This is the action of toric duality on the master space.

We interpret this result as a consequence (or a check) of Seiberg duality. Theories corresponding to
different phases are Seiberg dual for $N>1$ (here and henceforth $N$ denotes the number of D3 branes probing the singularity). Dual theories have the same spectrum of chiral operators, which can be organized into representations of the non-anomalous global symmetry group.  Anomalous symmetries, on the other hand, are not physical and can be different in different phases.  As argued in \cite{Forcella:2007wk,Forcella:2008bb,Butti:2007jv}, the generating function for the largest component of the moduli space of the theory with $N$ branes can be reconstructed from the knowledge of the Hilbert series for the coherent component of the master space. This is done by decomposing the Hilbert series into sectors with definite baryonic charge and by counting symmetric products through the Pletystic Exponential in each sector. It is then an interesting check of the procedure just explained that the Hilbert series for the coherent component of different phases is the same when expressed in terms of the non anomalous charges.

The paper is organized as follows. In Section 2 we discuss the general problem and present the main result of this paper in the form of a conjecture. We next proceed with examples. We analyze chiral and non-chiral theories. There is a class of toric non-chiral quiver gauge theories which is characterized by toric diagrams that have no internal points. This class is well studied in the literature and consists of the $\mathbb{C}^3/ ( \mathbb{Z}_2\times \mathbb{Z}_2 ) $ orbifold, together with the $L^{aba}$ class of theories, introduced in \cite{Uranga:1998vf,Erlich:1999rb} and further studied in \cite{Benvenuti:2005ja,Franco:2005sm,Butti:2005sw}. The $\mathbb{C}^3/ ( \mathbb{Z}_2\times \mathbb{Z}_2 ) $ model, being an orbifold of $\mathbb{C}^3$ has only one toric phase \footnote{Orbifolds of $\IC^3$ always have a single toric phase since the number of flavors per gauge group is always 3 and therefore any Seiberg duality will change the rank of the gauge group.} and therefore is trivial for the discussion at hand that requires toric singularities which have more than one toric phase. $L^{aba}$ theories have a number of toric phases and their master space is studied in Section 3. The number of toric phases depends on $a$ and $b$ and is computed in section \ref{numbertoric}. In Section 4, we first analyze in detail the case of the two phases of $\mathbb{F}_0$, which is the simplest chiral model with more than one phase, and for which we can also compute the generating functions for $N>1$.  We then analyze selected examples of chiral theories, including $dP_2$ and $dP_3$.
In Section 5, we discuss the generating functions for $N>1$.
The paper ends with conclusions and an Appendix on technical details about the Hilbert Series. 

\section{The Coherent Component}

We analyze the quiver gauge theories living on D3 branes at singularities focusing on toric phases where all gauge groups have the same rank, $N$, equal to the number of D3 branes. The gauge symmetry is thus $U(N)^g$, where $g$ is the number of gauge groups.

The master space  $\f{~}$ is defined as the set of solutions of the F-term constraints when all the chiral multiplets are regarded as c-numbers. Since all abelian groups decouple in the IR, this is the same as the IR moduli space of the quiver theory for $N=1$. The master space $\f$ is a toric variety of dimension $g+2$. The $g+2$ toric action corresponds to the global $U(1)$ symmetries of the quiver theory: three mesonic symmetries corresponding to the isometries of $X$ plus $g-1$ baryonic symmetries corresponding to the $g-1$ $U(1)$ gauge symmetries that decouple in the IR. There are $g-1$ gauge symmetries since the overall $U(1)$ in the quiver is decoupled from the Lagrangian. Only $d$ of the $U(1)$ symmetries are anomaly free, where $d$ is the number of external points in the toric diagram \footnote{$d$ counts the number of   integer points on the perimeter of the toric diagram; integer points on the sides of the diagram, which correspond to orbifold singularities of the
  base of $X$, should be counted.}. The three mesonic symmetries are always non-anomalous.

The master space is generically reducible. It decomposes into a large non-trivial component, called the coherent component and denoted $\firr{~}$, plus other lower dimensional pieces, typically linear. This is similar to the familiar decomposition of the moduli space of $\CN=2$ supersymmetric theories into Higgs and Coulomb branches. As shown in \cite{Forcella:2007wk,Forcella:2008bb}, $\firr{~}$ is a $g+2$ dimensional Calabi Yau cone. For $N=1$ there are no gauge groups and no strong gauge dynamics. The moduli space can be computed at the classical level. The Abelian nature does not allow the use of Seiberg duality to argue that the $N=1$ moduli space is the same for different phases. In fact, we will see that, in general, different phases have different $N=1$ moduli spaces. In particular, the structure and the number of linear components are different. More subtle is the fate of the coherent component under toric duality that we now examine. 

A very useful tool to characterize a toric variety is its  fully refined Hilbert series.  We introduce a set of auxiliary parameters (fugacities) $\{ t_i\}_{i=1}^{g+2}$ and define the generating function for holomorphic functions
\begin{equation}
g_1(\{ t_i\}) = \sum n_{k_1,\ldots,k_{g+2}} t_1^{k_1}\cdots t_{g+2}^{k_{g+2}} ,
\label{HS}\end{equation}
where $n_{k_1,\ldots,k_{g+2}}$ is the number of holomorphic functions with charge $\{k_1,\ldots,k_{g+2}\}$ under the global $U(1)^{g+2}$ symmetry. The set of holomorphic functions on the master space is just given by all polynomials in the chiral fields modulo the F-terms and the Hilbert series can be viewed as the generating function for the $N=1$ chiral ring that includes both mesonic and baryonic objects. We can write Hilbert series for the full master space and for its coherent component. The latter being irreducible, affine, and toric is completely specified by its set of holomorphic functions. The fully refined Hilbert series contains all information about the coherent component.

On an empirical basis, it was observed in \cite{Forcella:2008bb} that the coherent components of different toric phases are not equal.  A particular feature is their fully refined Hilbert series which are different. This fully refined Hilbert series contains however both anomalous and anomaly free charges. Since we are interested in applications to quantum field theory, a generating function which is graded by all symmetries, including the anomalous ones, is too much to require. Anomalous charges are not good quantum numbers in theories with $N>1$ and cannot be used in comparing different phases. Only global non-anomalous symmetries are invariant under Seiberg duality. Correspondingly, the number of chiral operators with equal non-anomalous quantum numbers agree in different phases, but nothing can be said about their anomalous charges. It then makes sense to consider a Hilbert series which is partially refined with respect to all the $d$ non-anomalous charges. We will see through examples that this partially refined Hilbert series is now an invariant under toric duality and we formulate the following
\vskip 0.3 cm 

\noindent{\bf CONJECTURE} $\,\,$ 
Toric phases of quiver toric gauge theories have the same Hilbert Series of the Coherent Component of the master space $\firr{~}$, refined in terms of all the fugacities of the global non anomalous $U(1)$ field theory charges.
\
\
\vskip 0.3 cm 

A special case of this conjecture is when the quiver theory is non-chiral. In all such cases all global $U(1)$ symmetries are anomaly free and the coherent components of the various toric phases are invariant under toric (Seiberg) duality. On the other hand, for chiral theories the coherent components of different toric phases are generically different. The different algebraic structure can be learned and is specified by their fully refined (including anomalous charges) Hilbert Series.

\section{Non-Chiral Theories}

Quiver gauge theories are non-chiral if every edge in the quiver has the same number of arrows in both directions. See Figure \ref{dcon} for simple examples of non-chiral quivers. The number of non anomalous $U(1)$ global symmetries is $g+2$; equal to the dimension of the coherent component of the master space $\firr{~}$. In this case our general 
{\bf Conjecture} implies that the coherent components of Seiberg dual gauge theories are isomorphic algebraic varieties.

We use the $L^{aba}$ theories as archetypal examples of non-chiral theories. They are particularly easy to analyze in the context of this paper since there are many toric phases, all of which have a simple brane interval realization.  For every  $b\ge a$, positive integers, we have a singularity $L^{aba}$ and a quiver gauge theory that can be realized in Type IIA with D4 branes on a circle with $b$ NS branes and $a$ NS$^\prime$ branes (see Figure \ref{quivLaba}). Toric phases differ by a rearrangements of NS and NS$^\prime$ branes on the circle. The number of all possible arrangements is counted in Section \ref{numbertoric}.

Before entering into technical details and discussing specific examples of different phases, we summarize the results of our analysis. Different phases have different linear components, corresponding in part to a variety of Coulomb branches. The coherent component of the master space for the $L^{aba}$ theories is instead the same for all toric phases. It is a $a+b+2$ dimensional manifold of complete intersection given by a collection of $2a+2b$ variables which are subject to $a+b-2$ quadratic equations. The corresponding Hilbert series in one variable can be simply written as
\beq
\label{Habt1}
H_{ab}(t) = \frac{(1-t^2)^{a+b-2} } { (1-t)^{2(a+b)} } = \frac{(1+t)^{a+b-2} } { (1-t)^{a+b+2} } ,
\eeq
where $t$ gives weight 1 for each generator and the relations are always quadratic. This can be further refined by realizing that the master space has two non-abelian hidden global symmetries which are induced by the structure of the toric diagram and are related to the two singularities $\IC^2/\IZ_a$ and $\IC^2/\IZ_b$. There are two hidden global symmetries $SU(a)$ and $SU(b)$ under which the fields transform as one copy of the fundamental of each and one copy of the antifundamental of each, giving all together $2a+2b$ fields. With the abelian charges the symmetry of the master space is $SU(a)\times SU(b)\times U(1)^4$. We can parametrize the 4 $U(1)$ charges by assigning a fugacity $t_i, i=1,\ldots 4$ to each of the 4 external points in the toric diagram. Under these charges each copy of the 4 multiplets above carries a different charge. The relations are singlets of the non-abelian groups. Getting all this information together, the Hilbert series takes the form

\beq
\label{Habt2}
H_{ab}(t_i ) = (1-t_1 t_2)^{a-1} (1-t_3 t_4)^{b-1} PE \left [ t_1 [1,0,\ldots 0]_a + t_2 [0,\ldots,0,1]_a+ t_3 [1,0,\ldots 0]_b + t_4 [0,\ldots,0,1]_b\right ] . \eeq

A special case is when $a=0$ and the $L^{0b0}$ singularity is the ${\cal N} = 2$ supersymmetric orbifold $\IC^{2}/\IZ_{b}\times \IC$ with its corresponding ${\cal N} = 2$ supersymmetric quiver. In this case the global symmetry of the master space reduces to $SU(b)\times U(1)^3$ and the $2b$ generators of the coherent component of the master space transform as one copy of the fundamental $[1,0\ldots, 0]$ representation and one copy of the anti-fundamental $[0\ldots, 0,1]$ representation of $SU(b)$. There are only 3 external points in the toric diagram and the 3 fugacities $t_{i}, i=1,2,3$ can again be assigned one per each of the external points. The Hilbert Series takes the form

\beq
\label{H0bt2}
H_{0b}(t_1, t_{2}, t_{3} ) = (1-t_1 t_2)^{b-1} PE \left [ t_1 [1,0,\ldots 0]_b + t_2 [0,\ldots,0,1]_b + t_3 \right ] . \eeq

This coincides with equation \eref{Habt1} with $a=0$, all non-abelian fugacities are set to 1, and all $t_{i} = t$.

\subsection{The double conifold}

\begin{figure}[ht!]
\begin{center}
  \includegraphics[scale=0.6]{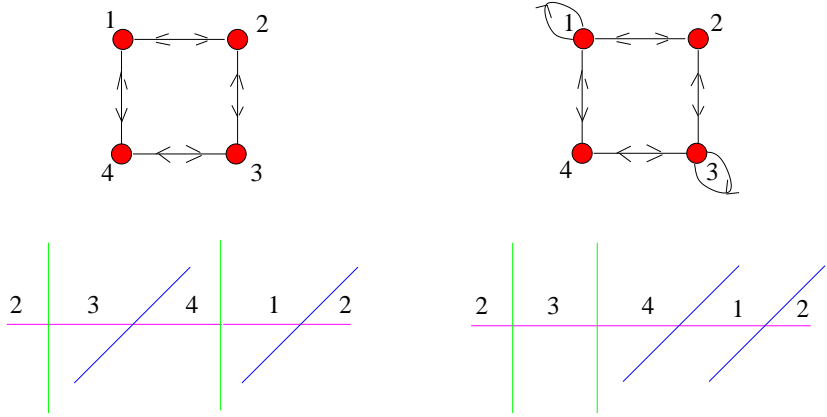}
\caption{\small The quivers for the two phases of the double conifold and the corresponding brane realizations in Type IIA in terms of D4 branes (in pink), NS5 branes (in green), and NS5$'$ branes (in blue). As is indicated in the picture Seiberg Duality is performed on gauge group 4.}
\label{dcon}
\end{center}
\end{figure}

The double conifold $L^{222}$ is the singular $\mathbb{Z}_2$ quotient of the conifold singularity. It is defined by the quadric $x^2y^2=wz$ in $\mathbb{C}^4$. The IR dynamics of a stack of $N$ regular D3 branes at the tip of the cone has two possible non-chiral UV descriptions. These are the two supersymmetric gauge theories with quivers given in Figure \ref{dcon} and superpotentials:
\begin{eqnarray}
& &W_I= X_{12} X_{21} X_{14} X_{41} - X_{21} X_{12} X_{23} X_{32} + X_{32} X_{23} X_{34} X_{43} - X_{43} X_{34} X_{41} X_{14} \nonumber\\
& &W_{II}= X_{11} (X_{12} X_{21} - X_{14} X_{41})  + X_{33}(X_{32} X_{23} - X_{34} X_{43}) + X_{43} X_{34} X_{41} X_{14} - X_{21} X_{12} X_{23} X_{32}\nonumber
\end{eqnarray}
These gauge theories have the same mesonic moduli space of vacua and they are related by toric (Seiberg) duality.  

The IR properties of Seiberg dual gauge theories must be equivalent, and it is interesting to understand what is the effect of Seiberg duality on the Master Space. For this purpose let us compute the Hilbert series for the two phases.

The Hilbert series for the master space for each of the two phases, written as a function of the fugacity for the $R$ charge, $t = t_{i}$ of equation \eref{Habt1} are
\begin{eqnarray}\label{Mducon}
& & H(t;(\f_{L^{222}})_I) = \frac{1 + 2 t + 3t^2 - 4t^3 + 2t^4}{(1 - t)^6} , \nonumber\\
& & H(t;(\f_{L^{222}})_{II})=\frac{(1 + 2t + 2t^2 - 2t^3 + t^4)^2}{(1 - t)^6(1 + t)^2} ,
\end{eqnarray}
where we give charge 1 to all bifundamental fields, and charge 2 to all adjoint fields \footnote{To match with R charge 2 for the superpotential we need to rescale by a factor of 1/2.}. The two Hilbert series are different. This fact gives a first hint that the two master spaces are indeed not isomorphic. A direct analysis of the algebraic equations defining the two Master Spaces reveals that they are not isomorphic varieties. They are indeed both reducible and given by: $(\f_{L^{222}})_I = (\firr{L^{222}})_I \cup (L^1_{L^{222}})_I \cup (L^2_{L^{222}})_I$, where:
\begin{eqnarray}\label{coII}
(\firr{L^{222}})_I &=& \mathbb{V}(X_{23} X_{32} - X_{41} X_{14}, X_{12} X_{21} - X_{34} X_{43}) \nonumber\\
(L^1_{L^{222}})_I  &=& \mathbb{V}(X_{14}, X_{41}, X_{32}, X_{23})\nonumber\\
(L^2_{L^{222}})_I &=& \mathbb{V}(X_{43}, X_{34}, X_{21}, X_{12}) 
\end{eqnarray}
and $(\f_{L^{222}})_{II} = (\firr{L^{222}})_{II} \cup (L^1_{L^{222}})_{II} \cup (L^2_{L^{222}})_{II} \cup (L^3_{L^{222}})_{II}$, where
\begin{eqnarray}\label{coI}
(\firr{L^{222}})_{II} &=& \mathbb{V}(X_{41}X_{14} - X_{33}, X_{34}X_{43} - X_{11}, X_{23} X_{32} - X_{11}, X_{12}X_{21} - X_{33}) \nonumber\\
 (L^1_{L^{222}})_{II} &=& \mathbb{V}(X_{43}, X_{34}, X_{32}, X_{23}, X_{11}, X_{12} X_{21} - X_{41} X_{14})\nonumber\\
 (L^2_{L^{222}})_{II} &=& \mathbb{V}(X_{14}, X_{41}, X_{21}, X_{12}, X_{33}, X_{23} X_{32} - X_{34} X_{43}) \nonumber \\
(L^3_{L^{222}})_{II} &=& \mathbb{V}(X_{14}, X_{41}, X_{43}, X_{34}, X_{32}, X_{23}, X_{21}, X_{12})
\end{eqnarray}
Here and in the following, $\mathbb{V}(\{ f_i(X)\})$ denotes the zero locus of the set of algebraic functions $f_i(X)$. 

Equations (\ref{coI}), (\ref{coII}) show two typical behaviors of the master space of non-chiral theories 
under Seiberg duality:
the master spaces $\f$ of different toric phases are in general non isomorphic and the number of smaller dimensional component $L^i$ is in general different; the coherent components $\firr{~}$ are instead isomorphic. In the case of the double conifold it is indeed easy to see that the coherent components defined by the equations in (\ref{coI}), (\ref{coII}) are isomorphic and define the product of two conifolds: 
\begin{equation} (\firr{L^{222}})_I=(\firr{L^{222}})_{II}= \mathcal{C}\times \mathcal{C} \, .\end{equation} 

One can also compute the Hilbert series of the coherent components of the two phases. They are obviously the same, equal to the Hilbert series of the product of two conifolds:

\begin{equation}\label{PPPP}
H(t;\firr{I})=H(t;\firr{II})=\frac{(1-t^2)}{(1-t)^4}\frac{(1-t^2)}{(1-t)^4} .
\end{equation}

This is consistent with equations \eref{Habt1} and \eref{Habt2}, setting $a=b=2, t_i = t$ and the non-abelian fugacities to 1. Moreover one can compute the fully refined Hilbert series for the two coherent components with all the six fugacities $t_i$, $i=1,\ldots,4$, for the abelian charges and $x_1, x_2$, for the non-abelian charges, all of which being non anomalous global symmetries: 
\bea\label{Hscompreft}
&&H(t_1,\ldots,t_4, x_1, x_2;\firr{I}) = H(t_1,\ldots,t_4, x_1, x_2;\firr{II}) = \\ \nonumber
&=& \frac{(1-t_1 t_2)(1-t_3 t_4)}{(1-t_1 x_1)(1-t_1/x_1)(1-t_2 x_1)(1-t_2/x_1)(1-t_3 x_2)(1-t_3/x_2)(1-t_4 x_2)(1-t_4 / x_2)} 
\eea
which is consistent with equation \eqref{Habt2}, with $x_1$ and $x_2$ weights for the $SU(2)\times SU(2)$ symmetry.
The full set of fugacities for the fields in the two phases of $L^{222}$ are given in  Table \ref{globalL222}.
\begin{table}[htdp]
\begin{center}
\begin{tabular}{|c|c|c|c|c|c|c|c|}
\hline
\ {\bf Phase I} \ & \ \  {\bf $t_1$}  \ \ & \ \  {\bf $t_2$}  \ \ & \ \  {\bf $t_3$} \ \ & \ \  {\bf $t_4$}  \ \ & \ \  {\bf $x_1$}  \ \ & \ \  {\bf $x_2$}  \ \ & \ \  fugacities \ \ \\
 {\bf fields}&  &  &  &  & &  & \ \  \ \  \\
\hline \hline
${\bf X_{12}}$  & $1$ & $0$ & $0$ & $0$ & $1$ & $0$ & $t_1 x_1$    \\
${\bf X_{34}}$  & $1$ & $0$ & $0$ & $0$ & $-1$ & $0$ & $t_1/x_1$  \\
${\bf X_{43}}$  & $0$ & $1$ & $0$ & $0$ & $1$ & $0$ & $t_2 x_1$   \\
${\bf X_{21}}$  & $0$ & $1$ & $0$ & $0$ & $-1$ & $0$ & $t_2/x_1$   \\
${\bf  X_{23}}$ & $0$ & $0$ & $1$ & $0$ & $0$ & $1$ & $t_3 x_2$  \\
${\bf  X_{41}}$ & $0$ & $0$ & $1$ & $0$ & $0$ & $-1$ & $t_3/x_2$  \\
${\bf X_{14}}$  & $0$ & $0$ & $0$ & $1$ & $0$ & $1$ & $t_4 x_2$  \\
${\bf X_{32}}$  & $0$ & $0$ & $0$ & $1$ & $0$ & $-1$ & $t_4/x_2$   \\
\hline \hline 
\ {\bf Phase II} \ & \ \  {\bf $t_1$}  \ \ & \ \  {\bf $t_2$}  \ \ & \ \  {\bf $t_3$} \ \ & \ \  {\bf $t_4$}  \ \ & \ \  {\bf $x_1$}  \ \ & \ \  {\bf $x_2$}  \ \ & \ \  fugacities  \ \ \\
{\bf fields}  &  &  &  &  & &  & \ \    \ \  \\
\hline \hline
${\bf X_{12}}$  & $1$  & $0$  & $0$ & $0$ & $1$ & $0$ & $t_1 x_1$    \\
${\bf X_{41}}$  & $1$  & $0$  & $0$ & $0$ & $-1$ & $0$ & $t_1/x_1$    \\
${\bf X_{33}}$  & $1$  & $1$  & $0$ & $0$ & $0$ & $0$ & $t_1 t_2$    \\
${\bf X_{14}}$  & $0$  & $1$  & $0$ & $0$ & $1$ & $0$ & $t_2 x_1$    \\
${\bf X_{21}}$  & $0$  & $1$  & $0$ & $0$ & $-1$ & $0$ & $t_2/x_1$    \\
${\bf X_{23}}$  & $0$  & $0$  & $1$ & $0$ & $0$ & $1$ & $t_3 x_2$    \\
${\bf X_{34}}$  & $0$  & $0$  & $1$ & $0$ & $0$ & $-1$ & $t_3/x_2$    \\
${\bf X_{11}}$  & $0$  & $0$  & $1$ & $1$ & $0$ & $0$ & $t_3 t_4$    \\
${\bf X_{43}}$  & $0$  & $0$  & $0$ & $1$ & $0$ & $1$ & $t_4 x_2$    \\
${\bf X_{32}}$  & $0$  & $0$  & $0$ & $1$ & $0$ & $-1$ & $t_4/x_2$    \\
\hline
\end{tabular}
\end{center}
\caption{Global charges for the basic fields of the two phases of the quiver gauge theory living on the D-brane probing the CY with $L^{222}$ base. Phase II is computed from phase I by dualizing node 4.}
\label{globalL222}
\end{table}
For a $d$ dimensional toric variety the fully refined Hilbert series, with all the $d$ fugacities, associated with the $U(1)^d$ toric action, contains information that is in one to one correspondence with the coordinate ring of holomorphic functions of the toric variety. Hence also a correspondence to the points in the dual cone $\sigma^*$, modulo $SL(d,\mathbb{Z})$ transformations. This means that the fully refined Hilbert series defines the algebraic variety modulo isomorphisms ( see Appendix A for more details). Indeed equation (\ref{Hscompreft}) implies that $(\firr{L^{222}})_I$ and $(\firr{L^{222}})_{II}$ are isomorphic algebraic varieties, as we already deduced from the direct analysis of the algebraic equations.

\subsection{$L^{aba}$}

$L^{aba}$ with $b \ge a$ is an infinite class of non isolated singularities that includes the double conifold, previously analized, as a special case: $L^{222}$. The $L^{aba}$ singularities are described by the quadric $x^a y^b = w z$ in $\mathbb{C}^4$. It reduces to the equation for the double conifold for the particular values $a=b=2$. The $L^{aba}$ singularities contain two lines of non isolated singularities passing through the tip of the cone: $\mathbb{C}^2/\mathbb{Z}_a$ and $\mathbb{C}^2/\mathbb{Z}_b$.  

\subsubsection{The number of Toric phases for $L^{aba}$}\label{numbertoric}
With the help of the Polya's Enumeration Theorem we can count the number of toric phases for $L^{aba}$. The formula does not have an explicit expression but we can write a generating function which can compute the number of phases for given values of $a$ and $b$.
To start the counting we use the Type IIA brane realization of this set of theories with $b$ NS branes and $a$ NS$'$ branes. Toric phases differ by a different arrangement of these branes on the circle. We are thus led to count the number ways one can arrange $a$ objects of one type and $b$ objects of another type on a circle.
The problem has obviously a cyclic symmetry as we are ordering objects on a circle but in addition it has a dihedral symmetry as a reflection of these objects on the circle does not change the theory under discussion and the toric phase remains the same under the reflection.
We are thus led to use the enumeration theorem with the dihedral index.
A similar counting is done for the number of toric phases of the $Y^{pq}$ quivers in \cite{Hanany:2005hq, Benvenuti:2004wx}.
Define the Cyclic index to be
\beq
{\cal Z} (\IZ_p) = \frac{1}{p} \sum_{n|p} \varphi(n) x_n^{p/n},
\eeq
where $\varphi$ is the Euler Totient function defined by
\beq
\varphi(n) = n \prod_{p|n}  \left (1-\frac{1}{p} \right ) ,
\eeq
and the Dihedral index to be
\beq
{\cal Z} (\mathbb{D}_p) =
\begin{cases}
\frac{1}{2} {\cal Z} (\IZ_p) + \frac{1}{2} x_1 x_2^{(p-1)/2}, & p \quad {\rm odd} \\ \\
\frac{1}{2} {\cal Z} (\IZ_p) + \frac{1}{4} \left (x_2^{p/2} + x_1^2 x_2^{(p-2)/2} \right ), & p \quad {\rm even} .
\end{cases}
\eeq
The variables $x_n$ keep track of the objects with $n$ elements and it is enough to take for the case at hand
\beq
x_n = \lambda_1^n + \lambda_2^n.
\eeq
This means that for $n$ elements there can either be $n$ of one type or $n$ of the other.
Now comes the magic of Polya's theorem.
When evaluating ${\cal Z} (\mathbb{D}_{a+b})$ we find a homogeneous polynomial in 2 variables $\lambda_1$ and $\lambda_2$,
\beq
{\cal Z} (\mathbb{D}_{a+b}) (\lambda_1, \lambda_2) = \sum_{a,b} d_{a,b} \lambda_1^a \lambda_2^b ,
\eeq
and the desired result is the integer number $d_{a,b}$ which counts the number of toric phases for the $L^{aba}$ theories.
For amusement we list the first few cases
\bea
\nonumber
{\cal Z} (\mathbb{D}_{1}) &=& \lambda _1+\lambda _2, \\ \nonumber
{\cal Z} (\mathbb{D}_{2}) &=& \lambda _1^2+\lambda _2 \lambda _1+\lambda_2^2, \\ \nonumber
{\cal Z} (\mathbb{D}_{3}) &=& \lambda _1^3+\lambda _2 \lambda _1^2+\lambda _2^2 \lambda_1+\lambda _2^3, \\ \nonumber
{\cal Z} (\mathbb{D}_{4}) &=& \lambda _1^4+\lambda _2 \lambda _1^3+2 \lambda _2^2\lambda _1^2+\lambda _2^3 \lambda _1+\lambda _2^4, \\
{\cal Z} (\mathbb{D}_{5}) &=& \lambda _1^5+\lambda_2 \lambda _1^4+2 \lambda _2^2 \lambda _1^3+2 \lambda _2^3 \lambda_1^2+\lambda _2^4 \lambda _1+\lambda _2^5, \\ \nonumber
{\cal Z} (\mathbb{D}_{6}) &=& \lambda _1^6+\lambda _2\lambda _1^5+3 \lambda _2^2 \lambda _1^4+3 \lambda _2^3 \lambda _1^3+3\lambda _2^4 \lambda _1^2+\lambda _2^5 \lambda _1+\lambda _2^6, \\ \nonumber
{\cal Z} (\mathbb{D}_{7}) &=& \lambda_1^7+\lambda _2 \lambda _1^6+3 \lambda _2^2 \lambda _1^5+4 \lambda_2^3 \lambda _1^4+4 \lambda _2^4 \lambda _1^3+3 \lambda _2^5 \lambda_1^2+\lambda _2^6 \lambda _1+\lambda _2^7, \\ \nonumber
{\cal Z} (\mathbb{D}_{8}) &=& \lambda _1^8+\lambda _2\lambda _1^7+4 \lambda _2^2 \lambda _1^6+5 \lambda _2^3 \lambda _1^5+8\lambda _2^4 \lambda _1^4+5 \lambda _2^5 \lambda _1^3+4 \lambda _2^6\lambda _1^2+\lambda _2^7 \lambda _1+\lambda _2^8, \\ \nonumber
{\cal Z} (\mathbb{D}_{9}) &=& \lambda _1^9+\lambda_2 \lambda _1^8+4 \lambda _2^2 \lambda _1^7+7 \lambda _2^3 \lambda_1^6+10 \lambda _2^4 \lambda _1^5+10 \lambda _2^5 \lambda _1^4+7\lambda _2^6 \lambda _1^3+4 \lambda _2^7 \lambda _1^2+\lambda _2^8\lambda _1+\lambda _2^9 . \nonumber
\eea
We recognize the two phases of $L^{222}$ as the coefficient of $\lambda _2^2\lambda _1^2$.

\begin{figure}[h!!!]
\begin{center}
\includegraphics[scale=0.55]{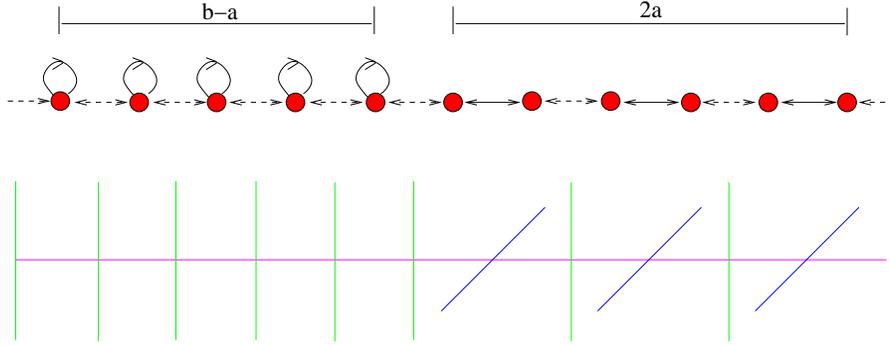}
\caption{\small The quivers for phase I of the $L^{aba}$ gauge theories, and their brane realizations in Type IIA}
\label{quivLaba}
\end{center}
\end{figure}

\subsubsection{Toric phases for $L^{aba}$ and Seiberg duality}\label{toricdualLaba}
The IR dynamics of $N$ D3 branes at the tip of the cone is a non-chiral gauge field theory with gauge group $\prod_{i=1}^{a+b}U(N)$. There 
are many different toric phases depending on the arrangement of the NS and NS$^\prime$ branes on the circle. These theories have different UV Lagrangians, with different field content and different superpotentials, but they are all equivalent in the IR and are related by toric (Seiberg) dualities.  

Let us denote by `phase I' the phase with the chiral field structure which is shown in the quiver in Figure \ref{quivLaba} and superpotential:
\begin{eqnarray}
\label{Labasup}
W=&&\sum_{i=1}^{b-a} X_{ii} (X_{i,i-1} X_{i-1,i}-X_{i,i+1}X_{i+1,i}) + \sum_{j=b-a+1}^{a+b}(-1)^{j+1}X_{j,j-1} X_{j-1,j} X_{j,j+1} X_{j+1,j} \nonumber\\
\end{eqnarray} 
where the index $i$ is cyclic modulo $a+b$ and the fields $X_{ii}$ transform in the adjoint representation of the $i$-th gauge group, while $X_{i,j}$ transforms in the fundamental representation of the $i$-th group and in the anti-fundamental of the $j$-th group.  We want to study the coherent component $\firr{L^{aba}}$ of the master space for this particular toric phase. The coherent component is by definition the locus of the F-flat term equations where generically each field has a non-zero vev. 

Let us start with the degenerate case $a=0$, $b=n$. In this particular case the gauge theories have $\mathcal{N}=2$ supersymmetry and $L^{0n0}=\mathbb{C}^2/\mathbb{Z}_n \times \mathbb{C}$. It is easy to show that $\firr{\mathbb{C}^2/\mathbb{Z}_n \times \mathbb{C}}=\f_{\mathbb{C}^2/\mathbb{Z}_n}\times \IC$, and
\begin{equation}\label{masterc2zn}
\f_{\mathbb{C}^2/\mathbb{Z}_n} = \mathbb{V}(X_{1,n} X_{n,1}-X_{1,2}X_{2,1},\ldots,X_{n,n-1} X_{n-1,n}-X_{n,1}X_{1,n}) .
\end{equation}
Namely $\firr{\mathbb{C}^2/\mathbb{Z}_n \times \mathbb{C}}$ is a product of equation (\ref{masterc2zn}) and the complex line parametrized by the adjoint fields, which are all equal. To simplify the discussion we will ignore the adjoint fields and the complex line $\IC$ in the 
geometry transverse to the D3 branes. 
If we call $x_i$, $i=1,\ldots,n$, the monomial $X_{i,i+1}X_{i+1,i}$ made with  the fundamental fields at the right of every gauge group, the equations (\ref{masterc2zn}), defining the master space of $\mathbb{C}^2/\mathbb{Z}_n$, are $x_1=x_2=\cdots=x_{n-1}=x_n$.

Let us now consider the generic $L^{aba}$ case. Let us enumerate the monomials  $X_{i,i+1}X_{i+1,i}$ as follows: those  drawn in solid line in Figure \ref{quivLaba} are called $x_i$, $i=1,\ldots,a$;  the ones drawn with dotted line are called $y_j$, $j=1,\dots,b$; and the adjoint fields are called $\phi_k$, $k=1,\dots,b-a$. The coherent components of the master spaces of the $L^{aba}$ theories is the locus of the F-term equations where generically all the fields are different from zero. It is given by the equations: $x_1=x_2=\cdots=x_{a-1}=x_a$ and $y_1=y_2=\cdots=y_{b-1}=y_b=\phi_1=\phi_2=\cdots=\phi_{b-a-1}=\phi_{b-a}$ in $\mathbb{C}^{3b+a}$. These equations describe algebraic varieties isomorphic to the zero locus $x_1=x_2=\cdots=x_{a-1}=x_a$, $y_1=y_2=\cdots=y_{b-1}=y_b$ in $\mathbb{C}^{2b+2a}$. These are the equations describing the master space of $\mathbb{C}^2/\mathbb{Z}_a$ and $\mathbb{C}^2/\mathbb{Z}_b$ respectively, and:

\begin{equation}
\firr{L^{aba}}=\f_{\mathbb{C}^2/\mathbb{Z}_a}\times \f_{\mathbb{C}^2/\mathbb{Z}_b}
\end{equation}

\begin{figure}[h!!!]
\begin{center}
\includegraphics[scale=0.55]{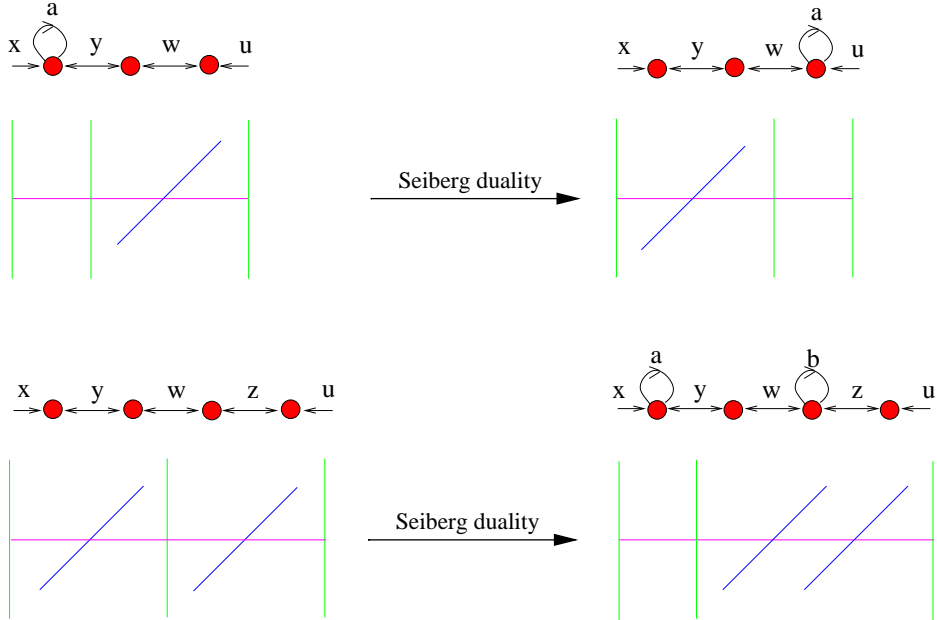}
\caption{\small Two relevant toric phases in $L^{aba}$ quiver gauge theories. The labels $x$, $y$, $w$, $z$, $u$, are for the 
quadratic monomials in the bifundamental fields, while the labels $a$, $b$ are for the adjoint fields.
}
\label{cohseib}
\end{center}
\end{figure}

As explained above, a given singularity generically corresponds to many UV field theories which are related by toric (Seiberg) dualities. The $L^{aba}$ gauge theories have an easy description in Type IIA in term of D4, NS5, NS5$'$ branes as shown in Figure \ref{quivLaba}. In this setup a Seiberg duality corresponds to the exchange of one NS5 brane with one NS5$'$ brane. Starting with the branes disposition in Figure \ref{quivLaba} there are only two exchanges of NS, NS$'$ branes that can affect the field theory content and the superpotential of the theory and they are shown in Figure \ref{cohseib}.

Seiberg duality is a local transformation of the quiver and of the superpotential and thus it is a local change in the F-term equations defining $\firr{L^{aba}}$. We will show that the local equations for the coherent component of the different toric phases in Figure \ref{cohseib} are algebraically equivalent.
Because the remaining part of the quiver is not changed by Seiberg duality the previous observation implies that Seiberg duality is an isomorphism of $\firr{L^{aba}}$. 

Let us use the labels defined in the Figure \ref{cohseib} for the quadratic monomials made with the bifundamental fields. There are two possible elementary steps.

The equations defining $\firr{~}$ for the upper left quiver are $x=y=u$ and $a=w$, while the equations defining the coherent component $\firr{S.d.}$ of the Seiberg dual quiver on the upper right are $x=w=u$ and $y=a$.  The two sets of equations are clearly isomorphic, hence $\firr{~}= \firr{S.d.}$.

Similarly, the equations defining $\firr{~}$ for the bottom left quiver  are $x=w=u$, and $y=z$, while the equations defining $\firr{S.d.}$ for the bottom right quiver  are $x=y=u=b$, and $w=z=a$. Once again these two sets of equations define isomorphic varieties, hence  $\firr{~}=\firr{S.d.}$.

All the possible Seiberg dual phases of $L^{aba}$ can be obtained by combining the two elementary transformations shown in Figure \ref{cohseib}. This means that the coherent component of the master space $\firr{L^{aba}}$ is invariant under Seiberg duality:

\begin{equation}
\firr{L^{aba}}=\firr{L^{aba}_{S.d.}}
\end{equation} 
Because $\firr{L^{aba}}$ is invariant under Seiberg duality the fully refined Hilbert series is invariant under Seiberg duality, and our general {\bf Conjecture} is right for the infinite class of the $L^{aba}$ gauge theories.

\section{Chiral theories}

Chiral theories 
are very common in the setup of D3 branes at singularities and in the AdS/CFT correspondence. For chiral theories part of the global $U(1)$ symmetries are anomalous and they do not have an explicit dual geometric interpretation. For this reason part of the $U(1)^{g+2}$ symmetries, which we use to completely characterize the toric varieties $\firr{~}$, is lost due to quantum dynamics. The dual geometric analysis done in \cite{Butti:2007jv,Forcella:2008bb} points towards the relevance of the coherent components of the master space $\firr{~}$ for the study of the complete moduli space and the BPS operators for $N>1$. In the following subsections we will do a case by case analysis to see what we can learn about the quantum dynamics using the concepts of the Master space and Hilbert series as tools of study.

\subsection{$\mathbb{F}_0$}

\begin{figure}[h!]
\begin{center}
  \epsfxsize = 11cm
  \centerline{\epsfbox{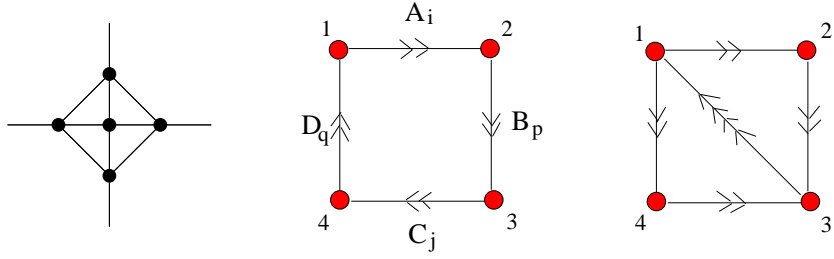}}
  \caption{{\sf The toric diagram and the quivers for phases I and II of $F_0$. Phase II is computed by dualizing node 4 of phase I.}}
  \label{f:F0III}
\end{center}
\end{figure}

The $\mathbb{F}_0$ theory is our first example of chiral gauge theory. It describes the low energy dynamics of a stack of D3 branes at the tip of the complex cone over $\mathbb{P}^{1} \times \mathbb{P}^{1}$. It has two toric phases \cite{Feng:2000mi, Feng:2001xr} with the quivers in Figure \ref{f:F0III}, and the superpotentials:

\begin{eqnarray}\label{WF0}
& & W_{\mathbb{F}_0^{I}} = \epsilon_{ij}\epsilon_{pq} {A}_i{B}_p{C}_j{D}_q \nonumber\\
& & W_{\mathbb{F}_0^{II}} = \epsilon_{ij}\epsilon_{mn} X^i_{12}X^m_{23}X_{31}^{jn}- \epsilon_{ij}\epsilon_{mn} X^m_{14}X^i_{43}X_{31}^{jn} \nonumber
\end{eqnarray}

The Master Spaces of the two phases, $\f_{\mathbb{F}_0^{I}}$, $\f_{\mathbb{F}_0^{II}}$, were computed in \cite{Forcella:2008bb},
\begin{eqnarray}
& &\f_{\mathbb{F}_0^{I}} = \firr{\mathbb{F}_0^{I}} \cup L^1_{\mathbb{F}_0^{I}} \cup L^2_{\mathbb{F}_0^{I}} , \nonumber\\
& &\f_{\mathbb{F}_0^{II}} = \firr{\mathbb{F}_0^{II}} \cup L^1_{\mathbb{F}_0^{II}} \cup L^2_{\mathbb{F}_0^{II}} \cup L^3_{\mathbb{F}_0^{II}} ,\nonumber
\end{eqnarray}
where the $L^i$ components are just copies of $\mathbb{C}^4$, while the two coherent components $\firr{~}$ are defined by the following equations in fields of the gauge theory:
\begin{eqnarray}\label{F0Ico}
& &\firr{\mathbb{F}_0^{I}}  = \mathbb{V}(B_2 D_1 - B_1 D_2, A_2 C_1 - A_1 C_2)
\end{eqnarray}
for the first phase, and:
\begin{eqnarray}\label{F0IIco}
\firr{\mathbb{F}_0^{II}}  &=& \mathbb{V}(X^2_{14} X^1_{23} - X^1_{14}X^2_{23}, X^2_{43}X^1_{12} - X^1_{43} X^2_{12}, X^1_{14}X^1_{43} - X^1_{12} X^1_{23}, X^2_{14} X^1_{43} - X^1_{12} X^2_{23},\nonumber \\ 
&&
X^1_{14} X^2_{43} - X^2_{12} X^1_{23}, X^2_{14} X^2_{43} - X^2_{12} X^2_{23}, X^{22}_{31} X^1_{23} - X^{21}_{31} X^2_{23}, X^{12}_{31} X^1_{23} - X^{11}_{31} X^2_{23},  \nonumber \\ 
&&
X^{22}_{31} X^1_{12} - X^{12}_{31} X^2_{12}, X^{21}_{31} X^1_{12} - X^{11}_{31} X^2_{12}, X^{12}_{31} X^2_{43} - X^{22}_{31} X^1_{43}, X^{11}_{31} X^2_{43} - X^{21}_{31} X^1_{43}, \nonumber \\
&&  X^{21}_{31} X^2_{14} - X^{22}_{31} X^1_{14}, X^{11}_{31} X^2_{14} - X^{12}_{31} X^1_{14}, X^{12}_{31} X^{21}_{31} - X^{11}_{31} X^{22}_{31}) \nonumber\\
\end{eqnarray}
for the second phase. 
From the equations (\ref{F0Ico}), (\ref{F0IIco}) it is easy to to see that $\firr{\mathbb{F}_0^{I}}$ is a complete intersection in 
$\mathbb{C}^8$ and it is isomorphic to the product of two conifolds: $\mathcal{C} \times \mathcal{C}$; while $\firr{\mathbb{F}_0^{II}}$ is a quite complicated not complete intersection in $\mathbb{C}^{12}$.

We would like to get an understanding on how different these two varieties are. The first step is to use a ``more toric'' description. Indeed the coherent component of the master space $\firr{~}$ is in general a toric, $g+2$ dimensional,
Calabi Yau cone, and it can be described by a symplectic quotient or a linear sigma model language. In \cite{Forcella:2008bb} it is shown that:
\begin{equation}
\firr{\mathbb{F}_0^{I}} \simeq \IC^8 // Q^t_I , \qquad \firr{\mathbb{F}_0^{II}} \simeq \IC^9 // Q^t_{II} ,
\end{equation}
where the charge matrices $Q$ are respectively:
\begin{equation}
Q^t_I = \tmat{ 0 & 0 & 0 & 0 & $-1$ & $-1$ & 1 & 1 
\cr 1 & 1 & $-1$ & $-1$ & 0 & 0 & 0 & 0 \cr  } , \qquad Q^t_{II} = \tmat{
 1 & 1 & 0 & -1 & 0 & -1 & 
    -1 & 0 & 1 \cr 0 & 1 & 0 & -1 & 0 & 0 & 
    -1 & 1 & 0 \cr 0 & 1 & -1 & 0 & 
    -1 & 1 & 0 & 0 & 0 \cr } .
\end{equation}
$\firr{\mathbb{F}_0^{I}}$ can be described by a toric diagram in $\mathbb{Z}_6$ (in this specific case $g+2=6$), with 8 vectors that satisfy 2 linear relations, while $\firr{\mathbb{F}_0^{II}}$ can be described by a toric diagram in $\mathbb{Z}_6$, with 9 vectors that satisfy 3 constraints. This means that the coherent components of the Master Spaces for the two phases of $\mathbb{F}_0$ are different toric varieties. Indeed the two $\firr{~}$ have two different toric diagrams that cannot be mapped one into the other by an $SL(6,\mathbb{Z})$ transformation. It is natural to wonder how much they are different. 

To answer this question let us compute some Hilbert Series as explained in the Appendix. 
The toric varieties $\firr{~}$ are $g+2$ dimensional and naturally admit $g+2$ fugacities associated to the imaginary $U(1)$ part of each $\IC^{*}$ in the $(\mathbb{C}^*)^{g+2}$ action. From the field theory point of view we can divide the fugacities into a set associated to the non anomalous symmetries and a set associated to the anomalous symmetries. Let us assign charges and fugacities to the elementary 
fields as in Table \ref{globalF0}.
\begin{table}[h]
\begin{center}
\begin{tabular}{|c|c|c|c|c|c|c|c|c|}
\hline
\ {\bf Phase I} \ & \ \  {\bf $F_1$}  \ \ & \ \  {\bf $F_2$}  \ \ & \ \  {\bf $R$} \ \ & \ \  {\bf $B$}  \ \ & \ \ {\bf $A_1$}  \ \ & \ \  {\bf $A_2$}  \ \ & \ \  fugacities   \\
\hline \hline
${\bf A}_1$  & $\frac{1}{2}$  & $0$      & $\frac{1}{2}$ & 1 & 1 & 0 & $t b x a_1$   \\
${\bf A}_2$  & $-\frac{1}{2}$ & $0$       & $\frac{1}{2}$ & 1 & 1 & 0 & $\frac{t b a_1}{x}$  \\
${\bf B}_1$  & $0$          & $\frac{1}{2}$ & $\frac{1}{2}$ &$-1$ & 0 & 1 & $\frac{t y a_2}{b}$  \\
${\bf B}_2$  & $0$          & $-\frac{1}{2}$ & $\frac{1}{2}$    &$-1$ & 0 & 1 & $\frac{t a_2}{b y}$   \\
${\bf C}_1$  & $\frac{1}{2}$    & $0$       & $\frac{1}{2}$ & 1 &$-1$ & 0 & $\frac{t b x}{a_1}$  \\
${\bf C}_2$  & $-\frac{1}{2}$   & $0$       & $\frac{1}{2}$ & 1 &$-1$ & 0 & $\frac{t b}{x a_1}$  \\
${\bf D}_1$  & $0$ & $\frac{1}{2}$ & $\frac{1}{2}$ &$-1$ & 0 &$-1$ & $\frac{t y}{b a_2}$   \\
${\bf D}_2$  & $0$          & $-\frac{1}{2}$    & $\frac{1}{2}$ &$-1$ & 0 &$-1$ & $\frac{t}{b y a_2}$ \\
\hline \hline
\ {\bf Phase II} \ & \ \  {\bf $F_1$}  \ \ & \ \  {\bf $F_2$}  \ \ & \ \  {\bf $R$} \ \ & \ \  {\bf $B$}  \ \ & \ \  {\bf $A_1$}  \ \ & \ \  {\bf $A_2$}  \ \ & \ \  fugacities   \\
\hline \hline
${\bf X^1_{12}}$ & $\frac{1}{2}$ & 0 & $\frac{1}{2}$ & 1 & $1 $ & $1 $ & $ t b x a_1 a_2$ \\
${\bf X^2_{12}}$ & $-\frac{1}{2}$ & 0 & $\frac{1}{2}$ & 1 & $1 $ & $1 $ & $ \frac{t b a_1 a_2}{ x} $ \\
${\bf X^1_{23}}$ & 0 & $\frac{1}{2}$ & $\frac{1}{2}$ & $-1$ & $0 $ & $-1 $ & $ \frac{t y}{b a_2}$ \\ 
${\bf X^2_{23}}$ & 0 & $-\frac{1}{2}$ & $\frac{1}{2}$ & $-1$ & $0 $ & $-1 $ & $ \frac{t}{b y a_2}$ \\  
${\bf X^{11}_{31}}$ & $\frac{1}{2}$ & $\frac{1}{2}$ & $1$ & $0$ & $-1 $ & $0 $ & $ \frac{t^2 x y}{a_1} $ \\ 
${\bf X^{12}_{31}}$ & $\frac{1}{2}$ & $-\frac{1}{2}$ & $1$ & $0$ & $-1 $ & $0 $ & $ \frac{ t^2 x}{y a_1}$ \\ 
${\bf X^{21}_{31}}$ & $-\frac{1}{2}$ & $\frac{1}{2}$ & $1$ & $0$ & $-1 $ & $0 $ & $ \frac{t^2 y}{x a_1}$ \\ 
${\bf X^{22}_{31}}$ & $-\frac{1}{2}$ & $-\frac{1}{2}$ & $1$ & $0$ & $-1 $ & $0 $ & $ \frac{t^2}{x y a_1}$ \\  
${\bf X^2_{14}}$ & $0$ & $-\frac{1}{2}$ & $\frac{1}{2}$ & $-1$ & $0$ & $1$ & $ \frac{t a_2}{b y}$ \\  
${\bf X^1_{14}}$ & $0$ & $\frac{1}{2}$ & $\frac{1}{2}$ & $-1$ & $0$ & $1$ & $\frac{t y a_2}{b}$ \\  
${\bf X^2_{43}}$ & $-\frac{1}{2}$ & $0$ & $\frac{1}{2}$ & $1$ & $1$ & $-1$ & $\frac{t b a_1}{x a_2}$ \\
${\bf X^1_{43}}$ & $\frac{1}{2}$ & $0$ & $\frac{1}{2}$ & $1$ & $1$ & $-1$ & $\frac{t b x a_1}{a_2}$ \\  
\hline
\end{tabular}
\end{center}
\caption{Global charges for the basic fields of the two phases of the quiver gauge theory
living on the D-brane probing the CY with $\mathbb{F}_0$ base.}
\label{globalF0}
\end{table}
$F_1$, $F_2$ are the flavor symmetries, 
$R$ is the R-symmetry, $B$ is the baryonic symmetry and $A_1$, $A_2$ are two anomalous $U(1)$ symmetries.
We introduce a fugacity $t$ for the R-charge, $x,y$ for the flavor charges, $b$ for the non anomalous baryonic symmetry and $a_1,a_2$ for the anomalous ones with the normalization indicated in the table.
We can compute the completely refined Hilbert Series for the two phases obtaining the result:
\begin{eqnarray}\label{HSF0}
 & &  H(x,y,t,b,a_1,a_2;~\firr{\mathbb{F}_0^{I}})= \\
& & \frac{ (1 - \frac{t^2}{b^2}) (1 - b^2 t^2) }{(1 - \frac{b t}{a_1 x})(1 - \frac{a_1 b t}{x})(1 - \frac{b t x}{a_1})(1 - a_1 b t x)(1 - \frac{t}{a_2 b y})(1 - \frac{a_2 t}{b y})(1 - \frac{t y}{a_2 b})(1 - \frac{a_2 t y}{b})} ;\nonumber \\
\nonumber\\
& & H(x,y,t,b,a_1,a_2;~\firr{\mathbb{F}_0^{II}})=   \frac{ P(x,y,t,b,a_1,a_2)} {(1-\frac{t^2 x y}{a_1})(1- \frac{ t^2 x}{y a_1})(1- \frac{t^2 y}{x a_1})(1- \frac{t^2}{x y a_1})} \times \nonumber \\ 
& &  \frac{1}{(1-  t b x a_1 a_2)(1-\frac{t b a_1 a_2}{x})(1-\frac{t y}{b a_2})(1-  \frac{t }{b y a_2})(1- \frac{t b x a_1}{a_2})(1- \frac{t b a_1}{ x a_2})(1-\frac{t y a_2}{b})(1-\frac{t a_2}{b y}) } \nonumber
\end{eqnarray}
with $P(x,y,t,b,a_1,a_2)$ a polynomial in the fugacities. One can check that the two Hilbert Series are really different. Now the interesting observation is that if we restrict just to the non anomalous charges, namely we put $a_1=a_2=1$, the two Hilbert Series become exactly the same:
\begin{eqnarray}
 H(x,y,t,b;~\firr{\mathbb{F}_0^{I}})=  H(x,y,t,b;~\firr{\mathbb{F}_0^{II}})=
 \frac{ (1 - \frac{t^2}{b^2}) (1 - b^2 t^2) }{(1 - \frac{b t}{x})^2(1 - b t x)^2(1 - \frac{t}{b y})^2(1 - \frac{t y}{ b})^2}\nonumber
\end{eqnarray}
This verifies our {\bf Conjecture}. We learn that the coordinate rings of the two varieties are exactly the same if labeled just in terms of the non anomalous charges. This fact, in a sense, defines how similar the two varieties are. If the two toric varieties $\firr{\mathbb{F}_0^{I}}$, $\firr{\mathbb{F}_0^{II}}$ are isomorphic then there must exist an $SL(2,\mathbb{Z})$ transformation on the $a_1$, $a_2$ fugacities that maps the Hilbert Series for the two phases. To see if this is possible let us expand the two functions in powers of $t$ near $t=0$,
\begin{eqnarray}
H(x,y,t,b,a_1,a_2;~\firr{\mathbb{F}_0^{I}}) &=& 1+ t b \left(x+\frac{1}{x} \right) \left(a_1+\frac{1}{a_1} \right) +\frac{t}{b} \left( y+\frac{1}{y} \right) \left(a_2+\frac{1}{a_2} \right) + \ldots\nonumber\\
H(x,y,t,b,a_1,a_2;~\firr{\mathbb{F}_0^{II}}) &=& 1+ t b  a_1\left(x+\frac{1}{x} \right) \left(a_2+\frac{1}{a_2} \right)  + \frac{t}{b} \left( y+\frac{1}{y} \right)  \left(a_2+\frac{1}{a_2} \right) + \ldots
\nonumber
\end{eqnarray}
It is easy to realize that the two series are quite similar in terms of the non anomalous charges but differ for the anomalous charges $a_i$, and there is no $SL(2,\mathbb{Z})$ transformation that can map one series into the other.

\subsection{$dP_2$}

\begin{figure}[h]
\begin{center}
  \epsfxsize = 11cm
  \centerline{\epsfbox{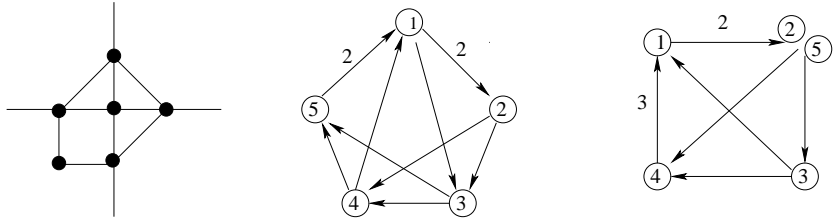}}
  \caption{{\sf The toric diagram and the quivers for phases I and II of $dP_2$. We use a block notation where numbers on the arrows denote number of fields between gauge groups. Phase II is computed from phase I by dualizing node 5.}}
  \label{dP2SD}
\end{center}
\end{figure}

The $dP_2$ theory is our second example of a chiral theory. It is the low energy gauge theory living on a stack of D3 branes at the tip of the complex cone over the second del Pezzo surface: $C_{\mathbb{C}}(dP_2)$. It has two toric phases \cite{Feng:2000mi, Feng:2001xr} with the matter content given by the quivers in Figure \ref{dP2SD} 
and superpotentials:
\begin{eqnarray}
W_I &=& X_{13} X_{34} X_{41} - Y_{12} X_{24} X_{41} + X_{12} X_{24} X_{45} Y_{51} - 
  X_{13} X_{35} Y_{51} \nonumber\\
& & + Y_{12} X_{23} X_{35} X_{51} - X_{12} X_{23} X_{34} X_{45} X_{51} ,\nonumber\\
\nonumber\\
W_{II} &=& Y_{41} X_{15} X_{54} - X_{31} X_{15} X_{53} + Y_{12} X_{23} X_{31} - Y_{12} X_{24} X_{41} + Y_{15} X_{53} X_{34} X_{41} \nonumber\\
& & - Z_{41} Y_{15} X_{54} + X_{12} X_{24} Z_{41} - X_{12} X_{23} X_{34} Y_{41} ,\nonumber\\
\end{eqnarray}
To analyze the Master Spaces and its coherent components we use the Hilbert Series. The master space is 7 dimensional and therefore we expect 7 $U(1)$ global symmetries, 4 of which are baryonic that further divide to 2 anomalous and 2 anomaly free charges. These last two $U(1)$ charges are expected to enhance to $E_2 = SU(2)\times U(1)$ \cite{Franco:2004rt}.  Let us refine the Hilbert Series with all the fugacities associated with the non-anomalous $U(1)$ global symmetries. We denote the 5 anomaly free fugacities by $t_i$,  as given in Table \ref{tabdP2}. Note that, in this notation, the symmetries $Q_i$ are all R charges.  
\begin{table}[htdp]
\begin{center}
\begin{tabular}{|c|c|c|c|c|c|c|c|c|}
\hline
{\bf Phase I} \ \ & \ \  {\bf $Q_1$}  \ \ & \ \  {\bf $Q_2$}  \ \ & \ \  {\bf $Q_3$} \ \ & \ \  {\bf $Q_4$}  \ \ & \ \ {\bf $Q_5$}\ \ & \ \ {\bf $A_1$}  \ \ & \ \  {\bf $A_2$}  \ \ & \ \  fugacities   \\
\hline \hline
${\bf X_{12}}$ & $1$ & $0$ & $0$ & $0$ & $0$  & $-2$ & $1$ & $t_1 a_2 / a_1^2$ \\
${\bf X_{35}}$ & $1$ & $0$ & $0$ & $0$ & $0$ & $-2$ & 1 & $t_1 a_2 / a_1^2$ \\
${\bf X_{41}}$ & $1$ & $1$ & $0$ & $0$ & $0$  & $1$ &  $1$  & $t_1 t_2 a_1 a_2$  \\ 
${\bf X_{23}}$ & $0$ & $1$ & $0$ & $0$ & $0$  & $3$ & $0$  & $t_2 a_1^3$ \\
${\bf Y_{51}}$ & $0$ & $1$ & $1$ & $0$ & $0$  & $1$ & $-2$ & $t_2 t_{3} a_1 / a_2^2$ \\
${\bf X_{34}}$ & $0$ & $0$ & $1$ & $0$ & $0$  & $-2$ & $-2$ & $t_3 / a_1^2 a_2^2$ \\
${\bf Y_{12}}$ & $0$ & $0$ & $1$ & $1$ & $0$  & $-2$ & $1$  & $t_3 t_{4} a_2 / a_1^2$ \\
${\bf X_{45}}$ & $0$ & $0$ & $0$ & $1$ & $0$  & $0$ & 3 & $t_4 a_2^3$ \\
${\bf X_{13}}$ & $0$ & $0$ & $0$ & $1$ & $1$  & $1$ & 1 & $t_4 t_5 a_1 a_2$ \\
${\bf X_{24}}$ & $0$ & $0$ & $0$ & $0$ & $1$  & $1$ & $-2$ & $t_5 a_1 / a_2^2$ \\
${\bf X_{51}}$ & $0$ & $0$ & $0$ & $0$ & $1$  & $1$  & $-2$ & $t_5 a_1 / a_2^2$ \\
\hline \hline
\ {\bf Phase II} \ & \ \  {\bf $Q_1$}  \ \ & \ \  {\bf $Q_2$}  \ \ & \ \  {\bf $Q_3$} \ \ & \ \  {\bf $Q_4$}  \ \ & \ \ {\bf $Q_5$}\ \ & \ \ {\bf $A_1$}  \ \ & \ \  {\bf $A_2$}  \ \ & \ \  fugacities \\
\hline \hline
${\bf X_{31}}$ & $1$ & $0$ & $0$ & $0$ & $1$  & $-1$ & $-1$ & $t_1 t_5 / a_1 a_2$ \\
${\bf X_{12}}$ & $1$ & $0$ & $0$ & $0$ & $0$  & $-2$ & $1$ & $t_1 a_2 / a_1^2$ \\
${\bf X_{54}}$ & $1$ & $0$ & $0$ & $0$ & $0$ & $1$ & $-2$ & $t_1 a_1 / a_2^2$ \\
${\bf X_{41}}$ & $1$ & $1$ & $0$ & $0$ & $0$  & $1$ &  $1$  & $t_1 t_2 a_1 a_2$  \\ 
${\bf X_{23}}$ & $0$ & $1$ & $0$ & $0$ & $0$  & $3$ & $0$ & $t_2 a_1^3$ \\
${\bf X_{15}}$ & $0$ & $1$ & $1$ & $0$ & $0$  & $-2$ & $1$ & $t_2 t_{3} a_2 / a_1^2$ \\
${\bf Z_{41}}$ & $0$ & $1$ & $1$ & $1$ & $0$  & $1$ &  $1$  & $t_{2} t_3 t_4 a_1 a_2$  \\ 
${\bf Y_{12}}$ & $0$ & $0$ & $1$ & $1$ & $0$  & $-2$ & $1$ & $t_3 t_{4} a_2 / a_1^2$ \\
${\bf X_{53}}$ & $0$ & $0$ & $0$ & $1$ & $0$  & $3$ & $0$ & $t_4 a_1^3$ \\
${\bf Y_{41}}$ & $0$ & $0$ & $0$ & $1$ & $1$  & $1$ &  $1$  & $t_4 t_5 a_1 a_2$  \\ 
${\bf X_{24}}$ & $0$ & $0$ & $0$ & $0$ & $1$  & $1$ & $-2$ & $t_5 a_1 / a_2^2$ \\
${\bf Y_{15}}$ & $0$ & $0$ & $0$ & $0$ & $1$  & $-2$ & $1$ & $t_5 a_2 / a_1^2$ \\
${\bf X_{34}}$ & $0$ & $0$ & $1$ & $0$ & $0$  & $-2$ & $-2$ & $t_3 / a_1^2 a_2^2$ \\
\hline
\end{tabular}
\end{center}
\caption{Global charges for the basic fields for the two phases of the quiver gauge theory living on the D-brane probing the CY with $dP_2$ base. Phase II is computed from phase I by dualizing node 5.}
\label{tabdP2}
\end{table}
The two baryonic charges $x, b$ of $E_2$ are related to the $t_i$'s by $t_1 = x/b, t_2 =  b^3/ x, t_3 = 1/b^4, t_4 = x b^3, t_5 =1/ bx.$ We summarize the translation between anomaly free charges in Table \ref{tabdP2a}.
\begin{table}[htdp]
\begin{center}
\begin{tabular}{|c||c|c|c|c|c|c|c|c|c|}
\hline
{Charges} \ \ & \ \ {\bf $R$}\ \ & \ \  {\bf $F_1$} \ \ & \ \  {\bf $F_2$}  \ \ & \ \  {\bf $SU(2)_H$}  \ \ & \ \  {\bf $B$}  \\
\hline \hline
 & & & & & \\
${Q_{1}}$ & $\frac{5\sqrt{33}-21}{16}$ & $\frac{1}{3}$ & $0$ & $1$ & $-1$ \\ 
 & & & & & \\
${Q_{2}}$ & $\frac{57-9\sqrt{33}}{16}$ & $0$ & $\frac{1}{2}$ & $-1$ & $3$ \\
 & & & & & \\
${Q_{3}}$ & $\frac{\sqrt{33}-5}{2}$ & $-\frac{2}{3}$ & $0$ & $0$ & $-4$ \\
 & & & & & \\
${Q_{4}}$ & $\frac{57-9\sqrt{33}}{16}$ & $0$ & $0$ & $1$ & $3$ \\
 & & & & & \\
${Q_{5}}$ & $\frac{5\sqrt{33}-21}{16}$ & $\frac{1}{3}$ & $-\frac{1}{2}$ & $-1$ & $-1$ \\
 & & & & & \\
\hline
\end{tabular}
\end{center}
\caption{A possible choice for the anomaly free baryonic and mesonic charges in terms of the 5 charges $Q_i$ of Table \ref{tabdP2} for the $dP_2$ theory. The baryonic symmetries can be enhanced to a non abelian symmetry $SU(2)_H\times B$  \cite{Franco:2004rt}; the role of this hidden symmetry, which does not commute with the flavor
symmetries, is still to be elucidated.}
\label{tabdP2a}
\end{table}

Let us start by computing the Hilbert Series for the complete Master Space $\f$ of the two phases in terms of just one anomaly free charge, obtained by setting all $t_i=t$. It is important to stress that $t$ is not a fugacity for the exact R symmetry, which is given in Table \ref{tabdP2a}, but rather it is a fugacity for a computationally convenient R-symmetry.
The Hilbert Series for the coherent component $\firr{~}$ turn out to be exactly the same in the two phases  \footnote{The Hilbert series for the coherent component can be computed using the matrix $K$ defined in \cite{Forcella:2008bb} which can be extracted from the F-term equations or, alternatively, from a symplectic quotient description of the master space using a Molien integral. We refer to \cite{Forcella:2008bb} for a detailed explanation of the computational techniques.}:
\begin{equation}
H(t;~\firr{(dP_2)_I}) = H(t;~\firr{(dP_2)_{II}}) =\frac{ 1 + 2 t + 5 t^2 + 2 t^3 + t^4}{(1 - t^2)^2(1 - t)^5} . \nonumber
\end{equation}
On the other hand one can easily check that the full Hilbert series for the master space is different, $H(t;~\f_{(dP_2)_I}) \not= H(t;~\f_{(dP_2)_{II}})$, meaning that $\f$ for the two phases is different and possibly reducible into different irreducible components. 

The Hilbert Series for $\firr{(dP_2)_I}$ and $\firr{(dP_2)_{II}}$, refined with the fugacities for all the non-anomalous symmetries are exactly the same:
\begin{eqnarray}
&& H(t_i;~\firr{(dP_2)_I}) = H(t_i;~\firr{(dP_2)_{II}}) \nonumber \\
 &&=   \frac{Q(t_i)}{(1-t_1)^2(1-t_2)(1-t_3)(1-t_4)(1-t_5)^2(1- t_1 t_2)(1-t_2 t_3)(1-t_3 t_4)(1-t_4t_5)}\nonumber
\end{eqnarray}
where $Q(t_i)$ is the palindromic polynomial:
\begin{eqnarray}
 Q(t_i) &=& 1 - ( t_1 t_2 t_3 + t_1 t_3 t_4 +t_1 t_4 t_5 +t_1 t_2 t_5 + t_2 t_3 t_5 + t_3 t_4 t_5)  \nonumber \\  & +&  (t_1 t_2 t_3 t_5 -t_1 t_2 t_3 t_4 - t_1 t_2 t_4 t_5 + t_1 t_3 t_4 t_5 - t_2 t_3 t_4 t_5) \nonumber \\
&+&  (t_1^2 t_2 t_3 t_4 + t_1^2 t_2 t_4 t_5 + t_1 t_2^2 t_3 t_5 + t_1 t_2 t_3^2 t_4 + t_2 t_3^2 t_4 t_5 + t_1 t_3 t_4^2 t_5 + t_1 t_2 t_4 t_5^2 + t_2 t_3 t_4t_5^2 + 4 t_1 t_2 t_3 t_4 t_5) \nonumber \\ &+& ( - t_1^2 t_2 t_3 t_4 t_5 + t_1 t_2^2 t_3 t_4 t_5 -t_1 t_2 t_3^3 t_4 t_5+ t_1 t_2 t_3 t_4^2 t_5 - t_1 t_2 t_3 t_4 t_5^2) \nonumber \\
 & -& (t_1^2 t_2^2 t_3 t_4 t_5 + t_1 t_2^2 t_3^2 t_4 t_5 +  t_1 t_2 t_3^2 t_4^2 t_5 + t_1 t_2 t_3 t_4^2 t_5^2 + t_1^2 t_2 t_3 t_4^2 t_5 + t_1 t_2^2 t_3 t_4 t_5^2 ) + t_1^2 t_2^2 t_3^3 t_4^2 t_5^2 \nonumber 
 \end{eqnarray}
thus confirming our general Conjecture.

We can also refine the Hilbert series with the remaining two fugacities $a_1$, $a_2$ associated to the field 
theory anomalous $U(1)$ symmetries. A 
computation using the charges in Table \ref{tabdP2} shows that the completely refined 
Hilbert series are different: $H(t_i,a_1,a_2;~\firr{(dP_2)_I}) \ne H(t_i,a_1,a_2;~\firr{(dP_2)_{II}})$.  

\begin{figure}[t]
\begin{center}
  \epsfxsize = 11cm
  \centerline{\epsfbox{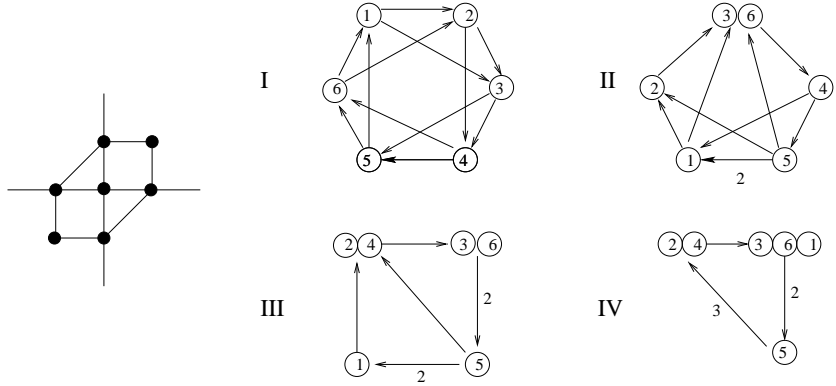}}
  \caption{{\sf The toric diagram and the quivers for phases I, II, III, IV of $dP_3$. We use a block notation where numbers on the arrows denote number of fields between gauge groups. Phases (II, III, IV) are computed from phases (I, II, III) by dualizing nodes (6, 4, 1) respectively.}}
  \label{dP3SD}
\end{center}
\end{figure}

\subsection{$dP_3$}

The $dP_3$ theory is our last example of a chiral theory. It is the low energy gauge theory living on a stack of D3 branes at the tip of the complex cone over the third del Pezzo surface: $C_{\mathbb{C}}(dP_3)$. It has four toric phases with the matter content given by the quivers in Figure \ref{dP3SD}  
and superpotentials \cite{Beasley:2001zp, Feng:2001bn}
\begin{eqnarray}
W_I &=& X_{13} X_{34} X_{46} X_{61} - X_{24} X_{46} X_{62} + X_{12} X_{24} X_{45} X_{51} - 
  X_{13} X_{35} X_{51} \nonumber\\
& & + X_{23} X_{35} X_{56} X_{62} - X_{12} X_{23} X_{34} X_{45} X_{56} X_{61} ,\nonumber\\
\end{eqnarray}

\begin{eqnarray}
W_{II} &=& X_{13} X_{34} X_{41} - X_{13} X_{35} X_{51} + X_{23} X_{35} X_{52} - X_{26} X_{65} X_{52} + X_{16} X_{65} Y_{51} \nonumber\\ 
& & - X_{16} X_{64} X_{41} + X_{12} X_{26} X_{64} X_{45} X_{51} - X_{12} X_{23} X_{34} X_{45} Y_{51} ,\nonumber\\
\end{eqnarray}

\begin{eqnarray}
W_{III} &=& X_{23} X_{35} X_{52} - X_{26} X_{65} X_{52} + X_{14} X_{46} X_{65} Y_{51} - X_{12} X_{23} Y_{35} Y_{51} + X_{43} Y_{35} X_{54} \nonumber\\ 
& & - Y_{65} X_{54} X_{46} + X_{12} X_{26} Y_{65} X_{51} - X_{14} X_{43} X_{35} X_{51} ,\nonumber\\
\end{eqnarray}

\begin{eqnarray}
W_{IV} &=& X_{23} X_{35} X_{52} - X_{52} X_{26} X_{65} + X_{65} Z_{54} X_{46} - Z_{54} X_{41} Y_{15} + Y_{15} Z_{52} X_{21} - Z_{52} X_{23} Y_{35} \nonumber\\
& & + Y_{35} X_{54} X_{43} - X_{54} X_{46} Y_{65} + Y_{65} Y_{52} X_{26} - Y_{52} X_{21} X_{15} + X_{15} Y_{54} X_{41} - Y_{54} X_{43} X_{35} . \nonumber\\
\end{eqnarray}

There are six non anomalous charges corresponding to the six external points of the toric diagram.
We use the assignment of charges given in \cite{Franco:2004rt} and reported in Table \ref{globaldP3}.
We use the fugacity $t$ to label the R symmetry and fugacities $t_{i}, i=1\ldots6$ for the 6 anomaly free
symmetries $Q_{i}, i=1\ldots6$. As for $dP_2$ the $Q_i$ are R-charges. A choice for mesonic and baryonic
anomaly free charges is reported in Table \ref{tabdP3a}.

\begin{table}[hhhh!!!!!!!!!!!!!!!!!!!!!!!!]
\begin{center}
\begin{tabular}{|c|c|c|c|c|c|c||c|c|c|c|c|c|c|}
\hline
\ {\bf I} \ & \ {\bf $Q_{1}$}  \  & \  {\bf $Q_{2}$}   \ & \  {\bf $Q_{3}$}  \ & \  {\bf $Q_{4}$}  \  & \   {\bf $Q_{5}$}  \  & \ \  {\bf $Q_{6}$}   \  & \ {\bf II} \ & \   {\bf $Q_{1}$}  \  & \   {\bf $Q_{2}$}  \ & \  {\bf $Q_{3}$}  \ & \ {\bf $Q_{4}$}  \ & \  {\bf $Q_{5}$}  \  & \  {\bf $Q_{6}$}   \\
\hline \hline
${\bf X_{35}}$ & $1$ & $0$ & $0$ & $0$ & $0$ & $1$ &
${\bf X_{35}}$ & $1$ & $0$ & $0$ & $0$ & $0$ & $1$ \\

${\bf X_{12}}$ & $1$ & $0$ & $0$ & $0$ & $0$ & $0$ &
${\bf X_{12}}$ & $1$ & $0$ & $0$ & $0$ & $0$ & $0$ \\

${\bf X_{46}}$ & $1$ & $1$ & $0$ & $0$ & $0$ & $0$ &
${\bf X_{41}}$ & $1$ & $1$ & $0$ & $0$ & $0$ & $1$ \\

${\bf X_{23}}$ & $0$ & $1$ & $0$ & $0$ & $0$ & $0$ &
${\bf X_{65}}$ & $1$ & $1$ & $0$ & $0$ & $0$ & $0$ \\

${\bf X_{51}}$ & $0$ & $1$ & $1$ & $0$ & $0$ & $0$ &
${\bf X_{23}}$ & $0$ & $1$ & $0$ & $0$ & $0$ & $0$ \\

${\bf X_{34}}$ & $0$ & $0$ & $1$ & $0$ & $0$ & $0$ &
${\bf X_{51}}$ & $0$ & $1$ & $1$ & $0$ & $0$ & $0$ \\

${\bf X_{62}}$ & $0$ & $0$ & $1$ & $1$ & $0$ & $0$ &
${\bf X_{34}}$ & $0$ & $0$ & $1$ & $0$ & $0$ & $0$ \\

${\bf X_{45}}$ & $0$ & $0$ & $0$ & $1$ & $0$ & $0$ &
${\bf X_{16}}$ & $0$ & $0$ & $1$ & $1$ & $0$ & $0$ \\

${\bf X_{13}}$ & $0$ & $0$ & $0$ & $1$ & $1$ & $0$ &
${\bf X_{52}}$ & $0$ & $0$ & $1$ & $1$ & $1$ & $0$ \\

${\bf X_{56}}$ & $0$ & $0$ & $0$ & $0$ & $1$ & $0$ &
${\bf X_{13}}$ & $0$ & $0$ & $0$ & $1$ & $1$ & $0$ \\

${\bf X_{24}}$ & $0$ & $0$ & $0$ & $0$ & $1$ & $1$ &
${\bf X_{45}}$ & $0$ & $0$ & $0$ & $1$ & $0$ & $0$ \\

${\bf X_{61}}$ & $0$ & $0$ & $0$ & $0$ & $0$ & $1$ &
${\bf X_{64}}$ & $0$ & $0$ & $0$ & $0$ & $1$ & $0$ \\

&&&&&&&
${\bf Y_{51}}$ & $0$ & $0$ & $0$ & $0$ & $1$ & $1$ \\

&&&&&&&
${\bf X_{26}}$ & $0$ & $0$ & $0$ & $0$ & $0$ & $1$ \\

\hline \hline
\ {\bf  III} \ & \   {\bf $Q_{1}$}  \ & \  {\bf $Q_{2}$}   \ & \  {\bf $Q_{3}$}  \ & \  {\bf $Q_{4}$}  \ & \  {\bf $Q_{5}$}   \ & \  $Q_{6}$  \ & \ {\bf IV} \ & \   {\bf $Q_{1}$}  \  & \   {\bf $Q_{2}$}  \  & \   {\bf $Q_{3}$} \  & \  {\bf $Q_{4}$}  \  & \  {\bf $Q_{5}$}   \ & \ $Q_{6}$ \\
\hline \hline

${\bf X_{12}}$ & $1$ & $0$ & $0$ & $0$ & $0$ & $0$ &
${\bf X_{41}}$ & $1$ & $0$ & $0$ & $0$ & $0$ & $0$ \\

${\bf X_{35}}$ & $1$ & $0$ & $0$ & $0$ & $0$ & $1$ &
${\bf X_{35}}$ & $1$ & $0$ & $0$ & $0$ & $0$ & $1$ \\

${\bf X_{54}}$ & $1$ & $1$ & $0$ & $0$ & $0$ & $1$ &
${\bf Z_{52}}$ & $1$ & $0$ & $0$ & $0$ & $1$ & $1$ \\

${\bf X_{65}}$ & $1$ & $1$ & $0$ & $0$ & $0$ & $0$ &
${\bf X_{65}}$ & $1$ & $1$ & $0$ & $0$ & $0$ & $0$ \\

${\bf X_{23}}$ & $0$ & $1$ & $0$ & $0$ & $0$ & $0$ &
${\bf Y_{52}}$ & $1$ & $1$ & $1$ & $0$ & $0$ & $0$ \\

${\bf X_{51}}$ & $0$ & $1$ & $1$ & $0$ & $0$ & $0$ &
${\bf X_{54}}$ & $1$ & $1$ & $0$ & $0$ & $0$ & $1$ \\

${\bf X_{46}}$ & $0$ & $0$ & $1$ & $0$ & $0$ & $0$ &
${\bf X_{23}}$ & $0$ & $1$ & $0$ & $0$ & $0$ & $0$ \\

${\bf Y_{35}}$ & $0$ & $0$ & $1$ & $1$ & $0$ & $0$ &
${\bf Y_{15}}$ & $0$ & $1$ & $1$ & $0$ & $0$ & $0$ \\

${\bf X_{52}}$ & $0$ & $0$ & $1$ & $1$ & $1$ & $0$ &
${\bf Y_{54}}$ & $0$ & $1$ & $1$ & $1$ & $0$ & $0$ \\

${\bf Y_{65}}$ & $0$ & $0$ & $0$ & $1$ & $1$ & $0$ &
${\bf X_{46}}$ & $0$ & $0$ & $1$ & $0$ & $0$ & $0$ \\

${\bf X_{14}}$ & $0$ & $0$ & $0$ & $1$ & $0$ & $0$ &
${\bf Y_{35}}$ & $0$ & $0$ & $1$ & $1$ & $0$ & $0$ \\

${\bf X_{43}}$ & $0$ & $0$ & $0$ & $0$ & $1$ & $0$ &
${\bf X_{52}}$ & $0$ & $0$ & $1$ & $1$ & $1$ & $0$ \\

${\bf Y_{51}}$ & $0$ & $0$ & $0$ & $0$ & $1$ & $1$ &
${\bf X_{21}}$ & $0$ & $0$ & $0$ & $1$ & $0$ & $0$ \\

${\bf X_{26}}$ & $0$ & $0$ & $0$ & $0$ & $0$ & $1$ &
${\bf Y_{65}}$ & $0$ & $0$ & $0$ & $1$ & $1$ & $0$ \\

&&&&&&&
${\bf Z_{54}}$ & $0$ & $0$ & $0$ & $1$ & $1$ & $1$ \\

&&&&&&&
${\bf X_{43}}$ & $0$ & $0$ & $0$ & $0$ & $1$ & $0$ \\

&&&&&&&
${\bf X_{15}}$ & $0$ & $0$ & $0$ & $0$ & $1$ & $1$ \\

&&&&&&&
${\bf X_{26}}$ & $0$ & $0$ & $0$ & $0$ & $0$ & $1$ \\

\hline
\end{tabular}
\end{center}
\caption{Global charges for the basic fields for the four phases of the quiver gauge theory
living on the D-brane probing the CY with $dP_3$ base.}
\label{globaldP3}
\end{table}

\begin{table}[htdp]
\begin{center}
\begin{tabular}{|c||c|c|c|c|c|c|c|c|c|}
\hline
{Charges} \ \ & \ \ {\bf $R$}\ \ & \ \  {\bf $F_1$} \ \ & \ \  {\bf $F_2$}  \ \ & \ \ {\bf $SU(3)_H$}  \ \ & \ \  {\bf $SU(2)_H$} \\
\hline \hline
 & & & & & \\
${Q_{1}}$ & $\frac{1}{3}$ & $-\frac{1}{3}$ & $-\frac{1}{6}$ & $1$ \quad $0$ & $-1$ \\ 
 & & & & & \\
${Q_{2}}$ & $\frac{1}{3}$ & $-\frac{1}{6}$ & $-\frac{1}{3}$ & $-1$ \quad $1$ & $1$ \\
 & & & & & \\
${Q_{3}}$ & $\frac{1}{3}$ & $\frac{1}{6}$ & $-\frac{1}{6}$ & $0$ \quad $-1$ & $-1$ \\
 & & & & & \\
${Q_{4}}$ & $\frac{1}{3}$ & $\frac{1}{3}$ & $\frac{1}{6}$ & $1$ \quad $0$ & $1$ \\
 & & & & & \\
${Q_{5}}$ & $\frac{1}{3}$ & $\frac{1}{6}$ & $\frac{1}{3}$ & $-1$ \quad $1$ & $-1$ \\
 & & & & & \\
${Q_{6}}$ & $\frac{1}{3}$ & $-\frac{1}{6}$ & $\frac{1}{6}$ & $0$ \quad $-1$ & $1$ \\
 & & & & & \\
\hline
\end{tabular}
\end{center}
\caption{A possible choice for the anomaly free baryonic and mesonic charges in terms of the 6 charges $Q_i$ of Table \ref{globaldP3} for the $dP_3$ theory. The baryonic symmetries can be enhanced to a non abelian symmetry $SU(3)_H\times SU(2)_H$  \cite{Franco:2004rt}; the role of this hidden symmetry, which does not commute with the flavor
symmetries, is still to be elucidated.}
\label{tabdP3a}
\end{table}

For simplicity, we consider only the coherent component of the four phases. The Hilbert series can be computed with the methods explained in \cite{Forcella:2008bb}.
Refining with only one fugacity $t_{i} = t$ we find
\begin{equation}
H(t;~\firr{(dP_3)_I}) = H(t;~\firr{(dP_3)_{II}})=H(t;~\firr{(dP_3)_{III}}) = H(t;~\firr{(dP_3)_{IV}})=
\frac{ 1 + 4 t^2 + t^4}{ (1-t)^6 (1-t^2)^2} \nonumber
\end{equation}
We see that, as expected, the coherent components have the same Hilbert series. The same is true
for the Hilbert series refined with all the six non anomalous charges which is given for all phases by
\begin{eqnarray}
&& H(t_i;~\firr{(dP_3)}) = \nonumber\\
 &&\hskip -1 cm \frac{P(t_i)}{(1-t_1)(1-t_2)(1-t_3)(1-t_4)(1-t_5)(1-t_6)(1-t_1t_2)(1-t_2t_3)(1-t_3t_4)(1-t_4t_5)(1- t_5t_6)(1-t_6t_1)}\nonumber
\end{eqnarray}
where $P(t_i)$ is the palindromic polynomial 
\begin{eqnarray}
\hskip -0.3cm P(t_i) &=& 1 - (t_1 t_2 t_3 t_4 +t_1 t_2 t_4 t_5 + t_2 t_3 t_4 t_5 +t_1 t_2 t_3 t_6 + t_1 t_3 t_4 t_6 + t_1 t_2 t_5 t_6 + t_2 t_3 t_5 t_6 + t_1 t_4 t_5 t_6 + t_3 t_4 t_5 t_6)\nonumber\\
&+& (t_1 t_2^2 t_3 t_4 t_5 + t_1 t_2 t_3 t_4^2 t_5 + t_1^2 t_2 t_3 t_4 t_6 + t_1 t_2 t_3^2 t_4 t_6 + t_1 t_2^2 t_3 t_5 t_6+
t_1^2 t_2 t_4 t_5 t_6 + 4 t_1 t_2 t_3 t_4 t_5 t_6 \nonumber\\
&+ & t_2 t_3^2 t_4 t_5 t_6 + t_1 t_3 t_4^2 t_5 t_6 + t_1 t_2 t_4 t_5^2 t_6 +
t_2 t_3 t_4 t_5^2 t_6 + t_1 t_2 t_3 t_5 t_6^2+t_1 t_3 t_4 t_5 t_6^2) \nonumber\\
&-& (t_1^2 t_2^2 t_3 t_4 t_5 t_6+ t_1 t_2^2 t_3^2 t_4 t_5 t_6+ t_1^2 t_2 t_3 t_4^2 t_5 t_6 + t_1 t_2 t_3^2 t_4^2 t_5 t_6 +
t_1 t_2^2 t_3 t_4 t_5^2 t_6 + t_1 t_2 t_3 t_4^2 t_5^2 t_6 \nonumber\\
&+& t_1^2 t_2 t_3 t_4 t_5 t_6^2 + t_1 t_2 t_3^2 t_4 t_5 t_6^2 + t_1 t_2 t_3 t_4 t_5^2 t_6^2) + t_1^2 t_2^2 t_3^2 t_4^2 t_5^2 t_6^2 \nonumber
\end{eqnarray} 

The fully refined Hilbert series depending on eight fugacities is instead different for the various phases.

The Hilbert series simplifies if the flavor charges $F_1,F_2$ are neglected. The fields can be organized  into representations of a symmetry $SU(3)_H\times SU(2)_H$, with $y_1,y_2$ weights for $SU(3)$ and $x$ weight for $SU(2)$ as discussed in \cite{Franco:2004rt}. 
The Hilbert series can be written as
\begin{equation}
H(t,y_1,y_2,x;~\firr{(dP_3)}) = \left ( 1+ t^2 [0,1;0] - [1,0;0] t^4 -t^6\right ) {\rm PE} \left [ t [1,0;1] + t^2 [0,1;0]\right ]
\end{equation}
It is not clear whether this expression implies the existence of a hidden symmetry, since the series expansion in representations contains negative signs. Moreover the $SU(3)_H\times SU(2)_H$ symmetry does not commute with
the flavor symmetry. A different $SU(3)\times SU(2)$, enhancing also  anomalous symmetries, was used in \cite{Forcella:2008bb} to find a positive sign  expansion for the Hilbert series of $dP_3$ into irreducible representations.

\section{The partition function for $N>1$}

The plethystic program can be efficiently applied to the study of the coherent component of the moduli space  for $N>1$ \cite{Butti:2007jv}. The Hilbert series $g_N(t; X)$, counting the combined baryonic and mesonic gauge invariant operators parameterizing the coherent component at finite $N$, is obtained from the Hilbert series for $N=1$, which we computed in the previous sections. In the notation of the previous Sections $g_1(t; X)\equiv H(t; \firr{X})$. 
As shown in \cite{Butti:2007jv},  the plethystic program requires a decomposition of the  $N=1$ generating function into sectors of definite baryonic charge, to which the plethystic exponential is applied.

Since the quiver gauge theories corresponding to different toric phases are Seiberg dual, the validity of the Pletystic program requires that it commutes with toric duality. As is shown below this is indeed the case.

The general construction in \cite{Butti:2007jv} is based on a decomposition of the Hilbert series $g_1(t_i;X)$
refined with the non anomalous charges 
\begin{equation}\label{GKZ1}
g_1(t_i;X) = \sum\limits_{\beta_1,\ldots, \beta_K} m(\beta_1, \ldots, \beta_K) g_{1,\beta_1, \ldots, \beta_K}(t_i; X)
\end{equation} 
on the lattice (the GKZ fan) of an auxiliary toric variety, which is the space of K\"ahler parameters of the original toric threefold $X$. This variety  is of dimension $K= I - 3 + d$, where $I$ is the number of internal points and $d$ is the number of vertices of the toric diagram of $X$. The lattice can be parametrized with a set of integer K\"ahler parameters $\beta_1, \ldots, \beta_K$. The generating functions $g_{1,\beta_1, \ldots, \beta_K}$ are geometrical in nature and they can be computed using
the equivariant index theorem, as given in Equation (4.18) of \cite{Butti:2007jv}. $g_{1,\beta_1, \ldots, \beta_K}$ is given by
a monomial in the baryonic fugacities multiplied by a non trivial function in the mesonic fugacities.
$m(\beta_1, \ldots, \beta_K)$ are integer multiplicities. 
We will not enter in the details of this construction and we refer the reader
to \cite{Butti:2007jv}.
The important point for our ensuing
discussions is that the plethystic program
can be applied to the $N=1$ partition functions at each point of the GKZ fan 
in order to obtain the finite  $N$ generating function 
\begin{equation}\label{GKZN}
g(t_i;X) := \sum_{N=0}^\infty \nu^N g_N(t; X) = \sum\limits_{\beta_1,\ldots, \beta_K} m(\beta_1,\ldots, \beta_K) PE \left [\nu  g_{1,\beta_1,\ldots, \beta_K}(t; X)\right ] \ .
\end{equation}
It is a general conjecture that all $N=1$ generating functions for toric quivers can be decomposed as in (\ref{GKZ1}). We tested this conjecture for a series of selected models. We will refer in the following to all the models where the previous construction is applicable.

Note that the construction is manifestly independent of the toric phase. In fact, as described above, $g_1(t_i;X)$, refined with the non anomalous charges, is the same in all toric phases and $g_{1,\beta_1, \ldots, \beta_K}$ can be computed from the geometry of $X$ only. This ensures that the GKZ prescription, when applicable, commutes with toric duality.

The $N=1$ master spaces of different phases are in general different algebraic varieties, but the spectrum of BPS operators parameterizing the coherent component, written in terms of the non anomalous charges, is the same, both for $N=1$ and for arbitrary $N>1$.

\subsection{$\mathbb{F}_0$ and multiplicities}

We now consider the specific example of $\mathbb{F}_0$ where we can write quite explicit formulae and discuss the issue of multiplicity.  

\begin{figure}[h]
\begin{center}
\includegraphics[scale=0.55]{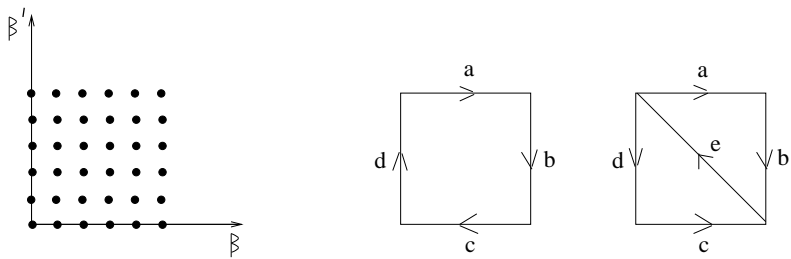}
  \caption{{\sf The Kahler GKZ decomposition for $\mathbb{F}_0$, and the GKZ quivers for phases $\mathbb{F}_0^{I}$ and $\mathbb{F}_0^{II}$.}}
\label{F0GKZ}
\end{center}
\end{figure}

The  K\"ahler moduli space for $\mathbb{F}_0$ is of dimension two. The localization partition functions  
$g_{1,\beta_1,\ldots,\beta_2}$ have been computed in \cite{Butti:2007jv} and read
\begin{eqnarray}
\label{ZBBF0}
g_{1,\beta_1,\beta_2} (t_1, t_2 , x, y; \mathbb{F}_0 ) &=& \frac{t_1^{\beta_1}t_2^{\beta_2} x^{-\beta_1} y^{-\beta_2}}{(1-x^2)(1- \frac{t_1^2 t_2^2}{x^2 y^2})(1-y^2)}+\frac{t_1^{\beta_1}t_2^{\beta_2} x^{\beta_1} y^{-\beta_2}}{(1-1/x^2)(1-t_1^2t_2^2x^2/y^2)(1-y^2)}\nonumber \\ 
&+&\frac{t_1^{\beta_1}t_2^{\beta_2} x^{-\beta_1} y^{\beta_2}}{(1-x^2)(1-t_1^2t_2^2y^2/x^2)(1-1/y^2)}+\frac{t_1^{\beta_1}t_2^{\beta_2} x^{\beta_1} y^{\beta_2}}{(1-1/x^2)(1-t_1^2t_2^2x^2y^2)(1-1/y^2)} \nonumber\\
\end{eqnarray}
where we set $t_1=t b, t_2= t/b$. Note that the dependence on $b$ of the functions $g_{1,\beta_1,\ldots,\beta_2}$ is purely multiplicative and is given by $b^{\beta_1-\beta_2}$. 

To use decomposition (\ref{GKZ1})  we need to compute the multiplicities $m(\beta_1, \ldots ,\beta_K)$ which in general is a difficult task. We proposed in \cite{Butti:2007jv} that, in simple models, the multiplicity of points in the GKZ lattice is counted by an {\bf auxiliary partition function}, so-called $Z_{\rm aux}$ and defined as follows. Take the simpler quiver than the original by neglecting any repeated arrows and then form the space of open but not closed loops in this simplified quiver. $Z_{\rm aux}$ is simply the (refined) Hilbert series of the ring of open paths modulo loops and relations and it is a generating function for multiplicities.  For example, in the case of $\mathbb{F}_0$, we can grade the ring with $t_1,t_2$ which play the role of GKZ parameters. Multiplicities can be read from the expansion
\begin{equation}
Z_{\rm aux}(t_1,t_2) = \sum\limits_{\beta_1,\beta_2} m(\beta_1, \ldots ,\beta_2) t_1^{\beta_1}t_2^{\beta_2}
\end{equation}
The procedure to determine the refined generating function $g_1$ in \eref{GKZ1} is now to replace a term $t_1^{\beta_1} t_2^{\beta_2}$ in $Z_{\rm aux}(t_1,t_2)$ by the expression for $g_{1,\beta_1,\beta_2}$.  

We now compute the auxiliary partition function for $\mathbb{F}_0$. In addition to $t_1,t_2$ we can use
the anomalous symmetries to grade the ring of open paths in the quiver. The auxiliary function depends on dimer combinatorics and, apparently, it depends on the toric phase. It is interesting to compute the fully refined  auxiliary function $Z_{\rm aux}(t_1,t_2,a_1,a_2)$  and compare the result obtained for the various toric phases.

In phase $I$ we can form a single loop in the simplified quiver, and the GKZ ideal is $abcd =0$. The auxiliary partition function is:

\begin{equation}
Z_{\rm aux}(t_1,t_2, a_1, a_2; \mathbb{F}_0^{I} ) =  \frac{1-t_1^2 t_2^{2}}{(1-t_1 a_1)(1-t_1/a_1)(1-t_2 a_2)(1-t_2/a_2)} .
\end{equation}
 
In phase II we have two closed loops  $abe, ced$ and two equivalent open paths, $ab$ and $ dc$. The GKZ ideal is $abe=0$, $ced=0$, $ab-dc=0$ and its partition function is

\begin{equation}\label{ZII}
Z_{\rm aux}(t_1,t_2, a_1, a_2; \mathbb{F}_0^{II} ) = \frac{(1- t_1 t_2 a_1)(1- t_1^2 t_2^2)}{(1-t_1 a_1 a_2)(1-t_1 a_1 / a_2)(1-t_2 a_2)(1-t_2 / a_2)(1- t_1 t_2/a_1)} .
\end{equation}
We see that the fully refined auxiliary partition functions are different in different phases. However, 
they become equal for $a_1=a_2=1$.  In particular the multiplicities $m_{\beta_1, \beta_2} \equiv m_{\beta_1, \beta_2}^{I, II}(a_1=1,a_2=1)$ do not depend on the phase.

The $N=1$  Hilbert series decomposition can be generalized to the fully refined $g_1(t_1,t_2,x,y,a_1,a_2)$. 
The two auxiliary partition functions can be expanded as
\begin{equation}
Z_{\rm aux}(t_1,t_2, a_1, a_2; \mathbb{F}_0^{I, II} ) = \sum\limits_{\beta_1,\beta_2} m_{\beta_1, \ldots ,\beta_2}^{I, II} (a_1,a_2) t_1^{\beta_1}t_2^{\beta_2}
\end{equation}
and we have the decomposition, valid in both phases,
\begin{equation}
g_1(t_1,t_2,x,y,a_1,a_2, \mathbb{F}_0^{I, II} ) = \sum_{\beta_1=0,\beta_2=0}^\infty m_{\beta_1, \ldots ,\beta_2}^{I, II} (a_1,a_2) g_{1,\beta,\beta^\prime}(t_1,t_2,x,y) .
\label{expa2F0}
\end{equation}

We see that the GKZ decomposition works in all toric phases of $\mathbb{F}_0$, with details that depend on the detailed form of the quiver when anomalous charges are introduced. Nevertheless, the Hilbert series $g_1(t_1,t_2,x,y;\mathbb{F}_0)$, the auxiliary partition function $Z_{\rm aux}(t_1,t_2; \mathbb{F}_0 )$ and the multiplicities $m_{\beta_1, \beta_2}$ do not depend on the toric phase. As a result, the partition function for the chiral ring for $N>1$ graded with non anomalous charges  is the same in all phases. This is consistent with Seiberg duality.

\section*{Acknowledgements}

We would like to thank J.~Davey for discussions.
D.~F.~is supported in part by INFN and the Marie Curie
fellowship under the programme EUROTHEPHY-2007-1.
A.~Z.~ is supported
in part by INFN and MIUR under contract 2007-5ATT78-002 and by the European Community's Human Potential Program MRTN-CT-2004-005104.

\section*{Appendix}
\appendix

\section{Toric geometry and Hilbert Series}

Consider an algebraic variety $\mathcal{V}$ defined as the zero locus of a set of polynomials $p_i \in\mathbb{C}[x_1,\dots,x_k]$ in $k$ variables. All regular algebraic functions on $\mathcal{V}$ are given by the restrictions of polynomials $\mathbb{C}[x_1,\dots,x_k]$ 
to $\mathcal{V}$.  The set of regular functions  has the algebraic structure of a ring, called the  coordinate ring of $\mathcal{V}$ and denoted by $\mathbb{C}[\mathcal{V}]$. It is  given explicitly by
\begin{equation}\label{coord}
\mathbb{C}[\mathcal{V}]=\frac{ \mathbb{C}[x_1,\dots,x_k]}{(p_1,\dots,p_l)} 
\end{equation}
The affine variety $\mathcal{V}$ is completely characterized by its coordinate ring, in the sense that two varieties
are isomorphic if and only if they have isomorphic coordinate rings \footnote{The variety, as a set of points, can be completely reconstructed from $\mathbb{C}[\mathcal{V}]$. In the more formal language of algebraic geometry, $\mathcal{V}$ is identified with the spectrum of the ring $\mathbb{C}[\mathcal{V}]$, i.e. the collection of its prime ideals, 
\begin{equation}\label{V}
\mathcal{V}=\hbox{Spec} \hbox{ }\mathbb{C}[\mathcal{V}] .
\end{equation}
}.
 
In this paper we make extensive use of the Hilbert Series (HS) to characterize the algebraic varieties we are interested in. Let us recall what a Hilbert Series is. Given an algebraic variety $\mathcal{V}$ with an action of an abelian group $U(1)^m$, we have an induced action on the coordinate ring $\mathbb{C}[\mathcal{V}]$.
Let us define a set of fugacities $t_i$, with $i=1,\dots,m$ for the action of $U(1)^m$.
The Hilbert Series is the generating function for the coordinate ring $\mathbb{C}[\mathcal{V}]$. It can be defined as the  
rational function whose expansion in  power series for small $t_i$ is
\beq\label{HSs}
H(t_1,\dots,t_m; \mathcal{V}) = \sum_{j_1,\dots,j_m} c_{j_1,\dots,j_m} t_1^{j_1}\cdots t_m^{j_m} \ ,
\eeq
where $c_{j_1,\dots,j_m}$ is the number of (algebraic) holomorphic functions  with $U(1)^m$ charges $j_1,\dots,j_m$. 

The Hilbert series contains a lot of information about the variety $\mathcal{V}$, but it is not in general a complete characterization of it. Different varieties may have the same Hilbert series. The story, however, is different in the context of toric geometry, which is the relevant case for this paper. After all the Master Space $\firr{~}$ for a three dimensional toric CY singularity $X$ is always a toric variety. As we discuss in this appendix toric, affine, irreducible varieties are completely determined by their fully refined Hilbert series (modulo a change of basis).

Recall that a toric variety $\mathcal{V}$ of dimension $n$ is a complex algebraic variety that admits an action of the $(\mathbb{C}^*)^n$ torus \cite{fulton}.
All the properties of an affine irreducible $n$ dimensional toric variety $\mathcal{V}$ and of its coordinate ring $\mathbb{C}[\mathcal{V}]$ are encoded in a set of combinatorial data,  the toric diagram $\sigma$, which is a rational polyhedral cone in $\mathbb{Z}^n$ defined by a set of integer vectors $V_{i}$, $i=1,\dots,d$. Equally important for the algebraic characterization of the variety is the dual cone 
$\sigma^*$ defined by, 
\beq
\label{cone}
{\cal \sigma}^* = \left \{ {\rm y}\in \mathbb{R}^n | l_{i} ( {\rm y}) = V^{j}_{i} y_j \ge 0, \, i = 1 \ldots d \right \} .\eeq 
The importance of $\sigma^*$ comes from the fact that  there is a one to one correspondence between integer points in $\sigma^*$ and  monomial functions $f(x_1,\dots,x_k)$ in the coordinate ring. There is exactly one monomial function for each point in the dual cone. This can be expressed at the algebraic level as 
\beq\label{Vtor}
\mathcal{V}  = \mbox{Spec} \hbox{  } \mathbb{C}[\mathcal{V}] = \mbox{Spec}[\sigma^*\cap \IZ^n] \ .
\eeq
The last equality in (\ref{Vtor}) means that the coordinate ring and the variety itself are completely determined 
by the dual cone $\sigma^*$.  

We can now define the HS as in (\ref{HSs}), and  refine it with as many fugacities $t_i$ as the complex dimension of the variety. To write the HS we need to compute charges for all the elements in the coordinate ring. This is particularly simple in the toric case, where the charges of a monomial function are given by the  integer coordinates of the corresponding point in $\sigma^*\cap \IZ^n$. In particular, there is a single holomorphic function with a specified set of charges $j_1,\dots,j_n$. This means that in Equation (\ref{HSs}), the coefficients $c_{j_1,\dots, j_n}$ have value $0$ or $1$. Hence the HS becomes a generating function for the integer points in the dual cone and it determines the variety itself.

The entire construction depends on a  choice of basis for $\mathbb{Z}^n$. All such choices are related by $SL(n,\mathbb{Z})$ transformations and give isomorphic varieties. 
Due to the freedom in choosing a basis for the lattice of charges, the functions have the same degree of arbitrariness. 
We conclude that, in the toric case, the fully refined HS with all the fugacities associated to the $n$ toric actions defines the toric variety up to $SL(n,\mathbb{Z})$ transformations.

For completeness, we describe the explicit algebraic-geometric description of the toric variety. We need to consider the cone $\sigma^*$ as a semi-group and find its generators over the integer numbers. The primitive vectors pointing along 
the edges generate the cone over the real numbers but we generically need to add other vectors to obtain a basis over the integers. We denote by $W_j$, with $j=1,\dots,k$, a set of generators of $\sigma^*$ over the integers.

\begin{equation}
\label{intdual}
 \sigma^* \cap \mathbb{Z} = Z_{\geq 0 } W_1 + \dots + Z_{\geq 0 } W_k
\end{equation}

The $k$ vectors $W_j$ generating the dual cone in $\mathbb{Z}^n$ are 
clearly linearly dependent  for $k > n$, and they satisfy some linear relations 

\begin{equation}
\label{reldual}
\sum _{j=1}^k p_{s,j} W_j = 0 \, \qquad\qquad p_{s,j}\in \mathbb{Z}
\end{equation}

To each vector $W_j$ we associate a coordinate $x_j$ in some 
ambient space $\mathbb{C}^k$.
The linear relations (\ref{reldual}) translate into a set of multiplicative relations 
among the coordinates $x_j$,
\begin{equation}
\label{cartacoord}
x_1^{p_{s,1}} x_2^{p_{s,2}}\cdots{  }x_k^{p_{s,k}} = 1\text{ for } \forall s   
\end{equation}
By clearing denominators, we obtain a set of polynomial equations for the affine toric variety   $\mathcal{V}$.

\subsection{A simple example}

We give an example of  two toric varieties which have the same Hilbert Series, when restricted to a particular  set of fugacities, but different fully refined  Hilbert Series. We use two familiar three dimensional Calabi Yau singularities: the conifold $\mathcal{C}$ and the $\mathbb{C}^2/\mathbb{Z}_2 \times \IC$ singularity. They can be embedded in $\mathbb{C}^4$, with coordinates $x,y,w,z$, using the two quadrics, respectively:
$$xy=wz \qquad, \qquad xy=w^2$$
These two singularities are clearly not isomorphic. Indeed if we define the charges of the coordinates as in Table \ref{conc2},
\begin{table}[htdp]
\begin{center}
\begin{tabular}{|c|c|c|c|c|c|c|c|c|}
\hline
$\mathcal{C}$ \ & \ \  $U(1)_1$  \ \ & \ \  $U(1)_2$  \ \ & \ \  $U(1)_3$ \ \ & \ \  $\mathbb{C}^2/\mathbb{Z}_2 \times \IC$ \ \ & \ \  $U(1)_1$  \ \ & \ \   $U(1)_2$  \ \ & \ \   $U(1)_3$ \\
\hline \hline
$x$  & 1  & 0 & 0  & $x$ &1  &0 &0  \\
$y$  & 0 & 1 & 0  & $y$ &0  & 1 & 0  \\
$w$  &0  & 0 & 1  & $w$ &1/2  & 1/2 & 0  \\
$z$  & 1 & 1 & $-1$  & $z$ & 0 & 0 &   1 \\
\hline
\end{tabular}
\end{center}
\caption{{\sf The mesonic charges of the Conifold ${\cal C}$ and of the orbifold $\IC^{2}/\IZ_{2}\times \IC$.}}
\label{conc2}
\end{table}
 and we introduce the fugacities $t_1$, $t_2$, $t_3$ for the three $U(1)$, the fully refined Hilbert Series for the two varieties are different
\begin{eqnarray}
 & & H(t_1,t_2,t_3; \mathcal{C})=\frac{1-t_1 t_2}{(1-t_1)(1-t_2)(1-t_3)(1-\frac{t_1 t_2}{t_3})} \nonumber\\
 & & H(t_1,t_2,t_3; \mathbb{C}^2/\mathbb{Z}_2 \times \IC )=\frac{1-t_1 t_2}{(1-t_1)(1-t_2)(1-\sqrt{t_1 t_2})(1- t_3)} \nonumber
\end{eqnarray}
and not related  by an $SL(3,\mathbb{Z})$ transformation.
It is easy to check that if we restrict to the diagonal $U(1)$ with fugacities $t_1=t_2=t_3=t$, the Hilbert Series become equal
\begin{equation}\label{HSconc2z2t}
H(t; \mathcal{C})=H(t; \mathbb{C}^2/\mathbb{Z}_2 \times \IC)= \frac{1-t^2}{(1-t)^4} \nonumber
\end{equation}
Here we see a situation similar to the ones studied in the main text: two different toric varieties, but with the same unrefined Hilbert Series.

In this specific case we can understand what is going on. The two toric varieties are related by a complex deformation.
Indeed they belong to the family of quadrics $xy= a wz + b w^2$, where $a$ and $b$ are two complex parameters, 
which interpolate between the conifold and the $A_1$ singularity. More generally, every quadratic equation in $(x,y,w,z)$  
has Hilbert Series 
\begin{equation}H(t;Q^2(\xi))= \frac{1-t^2}{(1-t)^4}\, .\end{equation} 
if we associate the fugacity $t$ to all the four variables $\xi = (x,y,w,z)$. 
From the unrefined Hilbert Series we learn that the conifold and the $A_1$ singularity belong to the same family of complex varieties, obtained by considering all quadrics  in $x$, $y$, $w$, $z$ that preserve the diagonal $U(1)$, but that generically break the other $U(1)$s.

\end{document}